\documentclass[12pt]{article}
\usepackage{pstricks,pst-node}
\usepackage{amsfonts}
\usepackage{sectsty}
\newcommand\ZZ{{\mathbb Z}}

\newcommand\CC{{\mathbb C}}
\newcommand\NN{{\cal N}}

\newcommand\RR{{\mathbb R}}
\newcommand\tr{\mathop\mathrm{tr}\nolimits}
\newcommand\Tr{\mathop\mathrm{Tr}\nolimits}
\newcommand\Det{\mathop\mathrm{Det}\nolimits}
\renewcommand\Im{\mathop\mathrm{Im}\nolimits}

\newcommand\eqdef{\buildrel {\rm def}\over=}
\renewcommand\int{\intop}
\renewcommand\oint{\ointop}
\newcommand\Z{{\cal Z}}
\newcommand\W{{\cal W}}
\newcommand\WW{\widetilde{\cal W}}
\newcommand\WH{\widehat{W}}
\newcommand\aq{\smash{\widetilde Q}\vphantom{Q}}
\newcommand\link{\Omega}
\newcommand\NH{\hat{N}}
\newcommand\FH{\hat{F}}
\newcommand\SH{\hat{S}}
\newcommand\RH{\hat{R}}
\newcommand\CH{\hat{C}}
\newcommand\EH{\hat{E}}
	
\newcommand\ws{{\textstyle{{\rm whole}\atop\Sigma}}}
\newcommand\vh{\hat{v}}
\newcommand\U{U}
\newcommand\M{{\mskip 2mu\cal U}}
\newcommand\A{A}
\newcommand\B{B}

\newcommand{\NS}[1]{\ensuremath{\frac{\hat{N}^{#1}}{\hat{S}^{#1}}}}
\newcommand{\SN}[1]{\ensuremath{\frac{\hat{S}^{#1}}{\hat{N}^{#1}}}}
\newcommand{\Prl}{\ensuremath{\prod_{\ell =1}^{K}}}
\newcommand{\Srl}{\ensuremath{\sum_{\ell =1}^{K}}}

\newcommand\ft{\ensuremath{^{\rm field}_{\rm theory}}}
\newcommand\mm{\ensuremath{^{\rm matrix}_{\rm model}}}

\newcommand\be{\begin{equation}}
\newcommand\ee{\end{equation}}
\newcommand\bea{\begin{eqnarray}}
\newcommand\eea{\end{eqnarray}}

\newcommand\vev[1]{\left\langle #1\right\rangle}

\newcommand\coeff[2]{{\textstyle{#1\over #2}}}
\newcommand\half{\coeff{1}{2}}

\newcommand\eq[1]{eq.~(\ref{#1})}
\newcommand\eqalign[1]{%
        \vcenter{%
                \normalbaselines \advance\baselineskip 5pt
                \advance\lineskip 5pt \tabskip=0pt
                \halign{%
                        &\hfil $\displaystyle{##{}}$&
                        $\displaystyle{{}##}$\hfil\cr
                        #1\crcr
                        }%
                }%
        }
\newcommand\eqrange[2]{eqs.~(\ref{#1}--\reftail{#2})}
\newcommand\dotline{\par\hbox to \hsize{\dotfill}\par}

\makeatletter
\def\befored@t#1.#2.#3;{#1}
\def\afterd@t#1.#2.#3;{#2}
\def\refhead#1{\edef\next{\ref{#1}}\expandafter\befored@t\next..;}
\def\reftail#1{\edef\next{\ref{#1}}\expandafter\afterd@t\next..;}
\def\lsim{\mathrel{\mathpalette\@versim<}}
\def\gsim{\mathrel{\mathpalette\@versim>}}
\def\@versim#1#2{\vcenter{\offinterlineskip
        \ialign{$\m@th#1\hfil##\hfil$\crcr#2\crcr\sim\crcr } }}
\newcommand\becomes[1]{\mathchoice{\becomes@\scriptstyle{#1}}
   {\becomes@\scriptstyle{#1}} {\becomes@\scriptscriptstyle{#1}}
   {\becomes@\scriptscriptstyle{#1}}}
\def\becomes@#1#2{\mathrel{\setbox0=\hbox{$\m@th #1{\,#2\,}$}%
        \mathop{\hbox to \wd0 {\rightarrowfill}}\limits_{#2}}}
\makeatother
\psset{unit=1mm,linecolor=black,linewidth=1pt}
\def\normalbaselines{%
        \normalbaselineskip=20pt plus 0.2pt minus 0.1pt
        \baselineskip=\normalbaselineskip
        \lineskip=2pt plus 0.1pt minus 0.1pt
        \lineskiplimit=2pt
        }
\def\fixformat{%
        \normalbaselines
        \abovedisplayskip=15pt plus 5pt minus 3pt
        \belowdisplayskip=\abovedisplayskip
        \parskip=6pt plus 2pt minus 1pt
        \skip\footins=20pt plus 10pt minus 5pt
        \predisplaypenalty=5000
        \postdisplaypenalty=500
        \interlinepenalty=50
        \interdisplaylinepenalty=10000
        \flushbottom
        }
\sectionfont{\huge\rm\blue\nohang\centering}
\subsectionfont{\Large\sc\nohang\raggedright}
\makeatletter
\def\@seccntformat#1{\@ifundefined{#1@cntformat}%
        {\csname the#1\endcsname\quad}
        {\csname #1@cntformat\endcsname}
        }
\def\section@cntformat{\thesection.\enspace}
\@addtoreset{equation}{section}
\newif\iffntmark \fntmarktrue
\renewcommand\@makefntext[1]{\noindent
        \iffntmark\llap{\@thefnmark\enspace}\fi
        #1\unskip }
\newif\ift@c \t@cfalse
\def\br@@k{\relax\ift@c\else\unskip\break\fi}
\def\brk{\protect\br@@k}
\let\t@c=\tableofcontents
\renewcommand\tableofcontents{\t@ctrue\t@c\t@cfalse}
\makeatother

\begin{document}
\bibliographystyle{hunsrt}
\fixformat
%
%
\begin{titlepage}

\rightline{\vbox{%
    \baselineskip=15pt \tabskip=0pt
    \halign{%
        #\unskip\hfil\cr
        UTTG--10--05\cr
        August 2005\cr
        hep-th/yymmxxx\cr
        }%
    }}
\vskip 2pc plus 1fil minus 0.5pc

\centerline{\large\bf Unitary Matrix Model of a Chiral $[SU(N)]^K$
Gauge Theory$\strut^\star$}
\vskip 2pc plus 1fil minus 0.5pc
\centerline{\large Edoardo Di Napoli$\strut^{\rm (a)}$ and
        Vadim S.\ Kaplunovsky$\strut^{\rm (b)}$}
\vskip 2pc plus 0.5pc minus 0.5pc
\tabskip=0pt plus 3in
\halign to \hsize{%
    \it #\hfil\cr
    Theory Group, Physics Department,\cr
    University of Texas at Austin,\cr
    Austin, TX~78712, USA\cr
    \ \rm (a) \tt edodin@physics.utexas.edu\cr
    \ \rm (b) \tt vadim@physics.utexas.edu\cr
    }
\tabskip=0pt
\vskip 2pc plus 1fil minus 0.5pc

\centerline{ABSTRACT}
\smallskip
\par\leftskip=20pt \rightskip=40pt
We build a matrix model of a chiral $[SU(N)]^K$ gauge theory ($\rm SQCD_5$
deconstructed down to 4D) using
random {\sl unitary} matrices to model chiral bifundamental
fields $({\bf N},{\bf\bar N})$ (without $({\bf\bar N},{\bf N})$).
We verify the duality by matching the loop equation of the matrix model
to the anomaly equations of the gauge theory.
Then we evaluate the matrix model's free energy and use it to derive
the effective superpotential for the gaugino condensates.
\par
\vskip 2pc plus 1fil minus 0.5pc

\begingroup
    \catcode`\@=11
    \let\@thefnmark=\relax
    \@footnotetext{%
        \nobreak
        \par\noindent \hangafter=1 \hangindent=20pt
	\rightskip 40pt
        {\large $\star$\enspace }
        Research supported in part by the National Science Foundation
        (grants PHY--0071512 and PHY--0455649) and by the 
        Office of Naval Research (Quantum Optics Initiative,
        grants N00014--03--1--0639 and N00014--04--1--0336.
        }
\endgroup

\end{titlepage}
\newpage
\pagenumbering{arabic}

\setcounter{page}{2}
%
%
\section{Introduction}
Three years ago, Robbert Dijkgraaf and Cumrun Vafa discovered a
peculiar duality between the $\NN=1$ supersymmetric gauge theories
in 4D and the bosonic matrix models without any spacetime at all.
At first~\cite{DiVa02Ma,DiVa02On}, they showed that the effective
superpotential for the gaugino condensates and the abelian gauge couplings
follow from planar diagrams in the gauge theory
(see also~\cite{DiGr02Pe,DiGu02Pe}), and then they argued~\cite{DiVa02Pe}
that the very same perturbative series also gives the free energy
of a bosonic random matrix model whose action is similar to the tree-level
superpotential of the gauge theory.
Shortly afterwards, Cachazo, Douglas, Seiberg, and Witten
\cite{CaDo02Ch,CaSe03Ph,CaSe03Ch} pointed out that the planar diagrams
describe the on-shell chiral ring of the gauge theory.
They used a different technique to study the on-shell ring, namely
the generalized Konishi anomaly equations, and those equations turned out
to be exactly similar to the loop equations of the random matrix model.
This confirmed the Dijkgraaf--Vafa gauge-matrix duality and made it more
precise:  the matrix model is dual to a subring of the gauge theory's
chiral ring comprising the gaugino condensates and the mesons;
other  operators  of the gauge theory are invisible to the matrix model.
In particular, the sphere-level free energy of the matrix model is dual to
the effective prepotential of the gaugino condensates.

Gauge-matrix duality works for all kinds of theories, including quiver
theories with multiple gauge groups and bifundamental matter~\cite{DiVa02On}.
Here is a brief summary of Dijkgraaf--Vafa rules for building the matrix model
of a particular gauge theory:
\par\begingroup
\parskip=0pt plus 2pt
\begin{itemize}
\item[{\bf1.}]
A $U(n)$ or $SU(n)$ gauge symmetry of the field theory becomes a $U(\NH)$
global symmetry of the matrix model, and we take the $\NH\to\infty$ limit.
If multiple gauge symmetries $SU(n_i)$ are involved, we take all
the $\NH_i\to\infty$ at the same rate: $\NH_i=t\times n_i$ for
the same  $t\to\infty$.
Similar rules apply to $SO(n)$ or $Sp(n)$ gauge group factors.
\item[{\bf2.}]
Chiral superfields becomes bosonic variables of the matrix model in similar
multiplets of the symmetry group.
Thus quark and antiquark superfields become complex vectors of length~$\NH$,
adjoint superfields become $\NH\times\NH$ matrices, and the bi-fundamental
multiplets of an $SU(n)\times SU(m)$ symmetry become
${\NH\times\hat M}$~matrices.
\item[{\bf3.}]
A {\it complex} chiral superfield becomes a {\it real}
bosonic variable of the matrix model (or rather $t$ or $t^2$ real variables.
Thus a complex adjoint superfield becomes a {\it hermitian}
${\NH\times\NH}$~matrix.
Likewise, a complex vector $A$ of the matrix model and its conjugate
$B=A^\dagger$ together represent both the quark and the antiquark superfields
of the gauge theory.
\par
\end{itemize}
\endgroup
\noindent
Rule {\bf3} works well for non-chiral gauge theories with real matter
multiples (or conjugate pairs like $\square+\overline{\square}$)
but not for chiral theories.
For example, consider an $SU(n)\times SU(m)$ theory with a {\sl chiral}
bifundamental field $({\bf n},\overline{\bf m})$ without the conjugate
$(\overline{\bf n},{\bf m})$ multiplets.\footnote{%
    The $({\bf n},\overline{\bf m})$ bifundamental must be accompanied
    by other chiral multiplets with opposite anomaly,
    but the present argument does not depend on those multiplets.
    }
By rules {\bf 1} and {\bf 2} this field becomes an $\NH\times\hat M$ matrix
transforming as a bifundamental of the $U(\NH)\times U(\hat M)$ symmetry,
and since this representation is complex, the matrix must be complex too.
But according to rule {\bf3} such 
complex $\NH\times\hat M$ matrix corresponds to
the non-chiral $({\bf n},\overline{\bf m})+(\overline{\bf n},{\bf m})$
multiplet of fields rather than the chiral $({\bf n},\overline{\bf m})$
without the~$(\overline{\bf n},{\bf m})$.

Lazaroiu {\it et~al.} \cite{Laza03Ho,LaLa03Ch}
found a way out of this problem:
instead of real bosonic variables integrated over~$\RR$, one may use
holomorphic variables, $i.\,e.$ complex variables integrated over
some contours in the complex plane rather than the whole plane~$\CC$.
More generally, one integrates $\NH\times\hat M$ matrix elements of
a complex matrix over some variety $\Gamma\subset\CC^{\NH\hat M}$
of {\sl real} dimension $\NH\hat M$.
The variety~$\Gamma$ must be consistent with the symmetries of the
matrix model according to rules~{\bf 1} and~{\bf2}.
For non-chiral models one may identify $\Gamma$ with the real ``axis''
$\RR^{\NH\hat M}$ of $\CC^{\NH\hat M}$ and recover rule~{\bf 3},
but chiral models require non-linear integration varieties.
Lazaroiu {\it et~al.} found such non-linear $\Gamma$ for the model of
a single--$SU(n)$ gauge theory with chiral
$\mathop{\square}\limits^{\vbox{\hbox{$\square$}\kern -3pt}}
+(n-4)\,\overline{\square}$ matter spectrum \cite{LaLa03Ch},
but in this article we interested in a different theory.

Specifically, we are interested in the chiral bifundamental fields,
hence we are looking for  varieties of $\NH\times\hat M$ complex matrices
which are invariant under the $U(\NH)_L\times U(\hat M)_R$ symmetry action.
To be precise, the variety~$\Gamma$ should satisfy
\be
\forall\, V_L\in U(\NH),\ \forall\, V_R\in U(\hat M):\quad
V_L^{}\Gamma V_R^\dagger\ \cong\ \Gamma
\label{eq:symmetry}
\ee
where `$\cong$' means equivalence as an integration variety:
same topology with respect to the singularities of the integrand, and similar
asymptotics when one or more matrix elements approach the complex infinity.
For square matrices $\NH\times\NH$ there is a simple solution, namely the
unitary group space $\Gamma=U(\NH)$ --- which is actually invariant under the
symmetry~(\ref{eq:symmetry}) ($V_L^{}\Gamma V_R^\dagger=\Gamma$)
and has the right real dimension${}=\NH^2$.
The problem is much more difficult for the rectangular matrices with
$\hat M\neq\NH$, and so we leave them for future research.
In this article, we work with the square matrices integrated over
$\Gamma=U(\NH)$.

On the gauge theory side, a unitary matrix $\U\in U(\NH)$ is dual to an
$({\bf n},{\bf\bar n})$ bifundamental field $\link$
with non-zero eigenvalues.
That is, the vacuum states of the field theory
which have matrix-model duals must have $\vev{\det\link}\neq 0$;
there may also be vacua with $\vev{\det\link}=0$, but the matrix model
might not work for such vacua.
Indeed, under the gauge-matrix duality, the allowed values of the fields
correspond to matrix variables belonging to the integration variety $\Gamma$
or any of its allowed deformations $\Gamma'\cong\Gamma$.
For example, a bosonic variable $z$ integrated over a unit circle with measure
$\oint\frac{dz}{z}$ can also be integrated over any other loop surrounding the
$z=0$ origin.
However, no such loop can go through the $z=0$ point itself because of the
measure singularity, and consequently
the field $\varphi$ dual to $z$ can take any complex values except $\varphi=0$.
Likewise, the unitary matrix integral can be deformed to integral over a variety
$\Gamma'\cong U(\NH)$ using an analytic extension
\be
\int_{\Gamma'}\!\!d\omega[\U]\
=\int_{\Gamma'}\!\!\prod_{\textstyle{(i,j)\atop\rm pairs}}(\U^{-1}dU)_{i,j}
\label{eq:Haar}
\ee
of the Haar measure.
However, this extension becomes singular for $\det\U=0$, which limits the
$\Gamma'$ integration varieties to the invertible matrices only.
Hence, on the gauge theory side of the duality we should have
$\det\link\neq0$, otherwise the duality might break down.

The bifundamental fields with $\det\link\neq0$ are common
in dimensional deconstruction.
Accordingly, in this article we build a unitary matrix model of an
$[SU(N_c)]^K$ 4D supersymmetric gauge theory which deconstructs
the 5D~SQCD \cite{DiKa05Qu};
the chiral ring of this theory was studied in great detail in~\cite{NaKa04Ch}.
The matter fields of the theory are shown on the following quiver diagram:
\par\allowbreak
\be
\psset{unit=4mm,linecolor=black}
\def\site{%
    \pscircle*[linecolor=green]{1}\relax
    \rput{*0}(0,0){\large $N_c$}\relax
    \psline{->}(-1.8,+0.45)(-0.9,+0.45)\relax
    \psline{->}(-1.8,-0.45)(-0.9,-0.45)\relax
    \psline{->}(-1.8,-0.15)(-0.99,-0.15)\relax
    \psline{->}(-1.8,+0.15)(-0.99,+0.15)\relax
    \psline{<-}(+1.8,+0.45)(+0.9,+0.45)\relax
    \psline{<-}(+1.8,-0.45)(+0.9,-0.45)\relax
    \psline{<-}(+1.8,-0.15)(+0.99,-0.15)\relax
    \psline{<-}(+1.8,+0.15)(+0.99,+0.15)\relax
    \rput(-2,0){\Large\{}\relax
    \rput(+2,0){\Large\}}\relax
    \rput{*0}(-2.9,0){$N_f$}\relax
    \rput{*0}(+2.9,0){$N_f$}\relax
    }
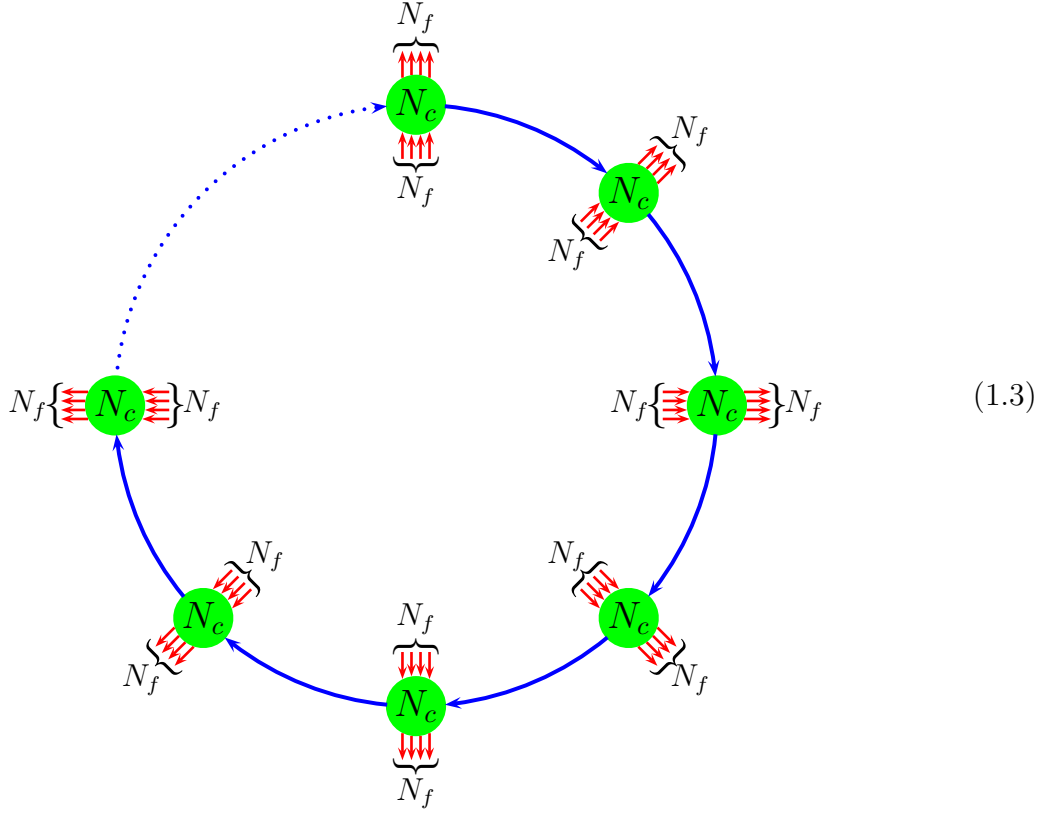
\begin{pspicture}[](-14,-14)(+14,+14)
    \psset{linewidth=1pt,linecolor=red}
    \rput{0}(+10,0){\site}
    \rput{45}(+7.07,+7.07){\site}
    \rput{315}(+7.07,-7.07){\site}
    \rput{90}(0,+10){\site}
    \rput{270}(0,-10){\site}
    \rput{225}(-7.07,-7.07){\site}
    \rput{180}(-10,0){\site}
    \psset{linewidth=1.5pt,linecolor=blue}
    \psarc{<-}{10}{50.5}{84.5}
    \psarc{<-}{10}{5.5}{39.5}
    \psarc{<-}{10}{320.5}{354.5}
    \psarc{<-}{10}{275.5}{309.5}
    \psarc{<-}{10}{230.5}{264.5}
    \psarc{<-}{10}{185.5}{219.5}
    \psarc[linestyle=dotted]{<-}{10}{95.5}{174.5}
\end{pspicture}
\label{fig:quiver}
\ee
In particular, the blue lines here denote chiral bifundamentals, which we shall
model via unitary matrices $\U_\ell$, $\ell=1,2,\ldots,K$.
The red lines denote the quark and the antiquark fields;
we shall model them in the usual way as complex $\NH\times\FH$ matrices
$\A_\ell$ and $\B_\ell^{}=A_\ell^\dagger$.

The article is organized as follows:
in the next section~\S2 we build the matrix model of
the gauge theory~(\ref{fig:quiver}) and establish
the gauge-matrix correspondence.
In particular, we derive the the loop equations of the matrix model and
see that they agree with the Konishi anomaly equations of the gauge theory.
In~\S3 we evaluate the matrix integral and derive the free energy of the matrix
model in terms of  contour integrals on the  spectral curve.
We begin by reducing the problem to an integral over a single unitary matrix,
and then we adapt the technology of
Dijkgraaf and Vafa~\cite{DiVa02Pe,DiVa02On,DiVa02Ma}
and Cachazo {\it et~al.}~\cite{CaDo02Ch} to the unitary case.
In~\S4 we derive the effective superpotential for the off-shell
gaugino condensates of the dual gauge theory.
Finally, in~\S5 we discuss open questions related to the present research.

 
%
%
\section{The Unitary Matrix Model\brk and its Loop Equations}
\subsection{The $[SU(N_c)]^K$ Gauge Theory and its Matrix Model}
Dimensional deconstruction of 5D~SQCD leads to an $\NN=1$
$[SU(N_c)]^K$ gauge theory in 4D as described in~\cite{IqKa04Qu,DiKa05Qu}.
The chiral ring of this $[SU(N_c)]^K$ theory was analyzed in much detail
in~\cite{NaKa04Ch}, and in this article we build and study its matrix model.
We begin with the basic structure of the 4D field theory as shown in the
quiver diagram~(\ref{fig:quiver}):
the green circles denote simple factors of the net gauge group
\be
G_{\rm 4D}\ =\  \prod_{\ell=1}^K [SU(N_c)]_{\,\ell}^{}
\ee
while the red and blue arrows denote the chiral superfields:
\be
\psset{unit=0.6mm,linecolor=red,linewidth=1pt}
\eqalign{
  \vcenter{\hbox to 9mm{%
        \psline{->}(0,6)(10,6)
        \psline{->}(0,4)(10,4)
        \psline{->}(0,2)(10,2)
        \psline{->}(0,0)(10,0)
        \hfil\Large\rm\}%
        }}\,
   \mbox{quarks}\ Q_{\ell,f}\ &
  =\ ( {\bf\square}_{\,\ell} ),\cr
  \vcenter{\hbox to 9mm{%
        \psline{<-}(0,6)(10,6)
        \psline{<-}(0,4)(10,4)
        \psline{<-}(0,2)(10,2)
        \psline{<-}(0,0)(10,0)
        \hfil\Large\rm\}%
        }}\,
    \mbox{antiquarks}\ \aq^f_\ell\ &
  =\ ( \overline{\bf\square}_{\,\ell} ),\cr
    \vcenter{\hbox to 25mm{%
        \psline[linecolor=blue,linewidth=1.5pt]{->}(0,0)(40,0)
        \hss }}\,
    \mbox{link~fields}\ \link^{}_\ell\ &
  =\ (\square_{\,\ell+1}, \overline\square_{\,\ell}),\cr
}\label{Fields}
\ee
where $f=1,2,\ldots,N_f$ and $\ell=1,2,\ldots,K$ is understood modulo $K$.%
\footnote{%
    For deconstruction purposes one takes the $K\to\infty$ limit, but from the
    4D point of view $K$ is a fixed parameter of the field theory;
    in our analysis we shall assume $K$ to be largish but finite.
    }
Note that the link fields form {\sl chiral} bifundamental multiplets
of the gauge group.

The tree level superpotential has three types of terms serving different
purposes,
\be
W_{\rm tree}\ =\ W_{\rm OR}\ +\ W_{\rm hop}\ +\ W_{\rm def}\,.
\ee
The O'Raifeartaigh terms
\be
W_{\rm OR}\ =\ \beta\sum_\ell s_\ell\times
\left(\det(\link_\ell)\,-\,v^{N_c}\right)
\label{eq:ORaifeartaigh}
\ee
--- where the $s_\ell$ are singlet fields not shown in the quiver diagram ---
turn each bifundamental field $\link_\ell$ into an $SL(N_c,\CC)$
linear sigma model.
This is important for deconstruction purposes, and also allows us to model
the $\link_\ell$ with unitary matrices without worrying about
the zero eigenvalues.
The hopping superpotential
\be
W_{\rm hop}\ =\ \gamma\Srl\sum_{f=1}^{N_f} \left(
        \aq^f_{\ell+1} \link^{}_\ell Q_{\ell,f}\
        -\ \mu_f^{}\, \aq^f_\ell Q_{\ell,f}
        \right)
\label{eq:hopping}
\ee
describes quarks' masses and interactions which let them `hop' between
quiver nodes; in 5D terms, this allows quark propagation in the
deconstructed $x^4$ direction.
%
%
Finally, we have the deformation superpotential
\be
W_{\rm def}\ =\ \tr\Bigl(\W(\link_K\link_{K-1}\cdots\link_2\link_1)\Bigr)
\equiv\,\sum_{p=1}^d\frac{g_p}{p}\tr
\Bigl((\link_K\cdots\link_1)^p\Bigr)
\label{eq:deformation}
\ee
for some polynomial $\W(X)=\sum_p\frac{g_p}{p}X^p$ with constant coefficients
$g_1,\ldots,g_d$.
The $W_{\rm def}$ deforms the 4D theory away from deconstructed $\rm SQCD_5$,
breaks the Coulomb branch of the moduli space into a discrete set of vacua,
and allows formation of the gaugino condensates.
This deformation is analogous to the tree-level superpotential for
the adjoint field in \cite{CaDo02Ch,Seib02Ad,CaSe03Ph} and is
essential for understanding the on-shell chiral ring of the theory.
It is also a key ingredient of the matrix model.

In the matrix model, the $[SU(N_c)]^K$ gauge symmetry of the field theory
becomes a $[U(\NH)]^K$ global symmetry and we take the
$\NH\to\infty$ limit.
Consequently, the bifundamental fields $\link_\ell$ become $\NH\times\NH$
unitary matrices $\U_\ell$, and for chirality's sake we should integrate
each matrix $\U_\ell$ over the $U(\NH)$ group manifold or an equivalent
variety.
However, in light of the O'Raifeartaigh terms~(\ref{eq:ORaifeartaigh})
we restrict the determinants of the $\U_\ell$ matrices and integrate them
over the $SU(\NH)$ group manifold.
To accommodate the $v^{N_c}$ factor, we rescale the field-matrix
correspondence according to $\link_\ell\leftrightarrow \vh\times\U_\ell$
where $\vh=v+{}$quantum corrections.
Such quantum corrections are computable in the field theory ---
{\it cf.}\ \S4.3 of \cite{NaKa04Ch} for details --- but in the
matrix model they need to be put in by hand.

The quark sector of the field theory is non-chiral --- for each quark~$Q_\ell$
there is an antiquark~$\aq_\ell$ with opposite quantum numbers.
In the matrix model, the quarks become complex rectangular $\NH\times\FH$
matrices~$\A_\ell$ while the antiquark fields $\aq_\ell$ become conjugate
$\FH\times\NH$~matrices $\B_\ell^{}=\A_\ell^\dagger$.
When we take the $\NH\to\infty$ limit, we have two options for the flavor
number~$\FH$ of the matrix model:
we may keep it fixed ($i.\,e.,\ \FH\equiv N_f$), or we may let it grow
while keeping the flavor/color ratio fixed,
$\FH/\NH\equiv N_f/N_c$~\cite{McGr02Ad}.
In the 't~Hooft limit of fixed $\FH$, the bifundamental sector dominates
the matrix model in the large color limit,
and the quark sector becomes quenched --- its backreaction on the
bifundamental sector becomes negligible.
This limit oversimplifies the physics but makes for a simple
$1/\NH$ perturbation theory in terms of Feynman-like diagrams' topology.
On the other hand, in the un-quenched limit of $\FH,\NH\to\infty$ the
matrix model has rich flavor physics, but the $1/\NH$ expansion becomes
much more difficult.
Consequently, we use the 't~Hooft limit in this article and leave the
un-quenched flavor physics for future research.

Together, the $\U_\ell$, $\A_\ell$, and $\B_\ell$ matrices comprise
the entire matrix model.
Its partition function is defined as the following matrix integral:
\be
\Z\ =\ {\cal C}\Prl\left\{\int_{SU(\NH)}\!\!d\omega[\U_\ell]
        \mathop{\int\!\!\!\int}\limits_{\B_\ell^{}=\A_\ell^\dagger}\!\!
        d\B_\ell\,d\A_\ell\right\}
\exp\left(-\NS\,\WH({\rm all}\ \U_\ell,\B_\ell,\A_\ell)\right)
\label{eq:Zfun}
\ee
where the matrix potential is
\be
\eqalign{
\WH(\U,\B,\A)\ ={}&
\WH_{\rm def}(\U)\ +\ \WH_{\rm hop}(\U,\B,\A)\cr
\noalign{\vskip 10pt}
{}=&\ \sum_{p=1}^d\frac{g_p \vh^{pK}}{p}\,
        \tr\Bigl((\U_K\U_{K-1}\cdots\U_2\U_1)^p\Bigr)\cr
&+\ \gamma\Srl\Bigl( \vh\tr(B_{\ell+1}\U_\ell\A_\ell)\
        -\ \tr(\hat\mu\A_\ell\B_\ell)\Bigr)\cr
}\label{eq:WH}\\
\ee
where $\hat\mu$ is an $\FH\times\FH$ matrix with eigenvalues
$\mu_1,\ldots,\mu_{\FH}$.
Note that the matrix potential does not contain O'Raifeartaigh terms analogous
to the $W_{\rm OR}$ of the field theory, but integrating the $\U_\ell$ over
the ${\bf S\rm U}(\NH)$ instead of ${\rm U}(\NH)$ has the same effect.

The denominator $\SH$ in the exponent in \eq{eq:Zfun} is the overall coupling
constant of the matrix model; as usual, $\SH$ is fixed while $\NH\to\infty$.
Under gauge-matrix duality, $\SH$ is dual to the net gaugino condensate
$S=\sum_i S_i$ of all subgroups of the $SU(N_c)_{\rm diag}=
\mathop{\rm diag}[{\rm all}\ SU(N_c)_\ell]$.
In the field theory, the $S_i$ and hence the $S$ emerge from the on-shell
chiral ring, and then need to be integrated in to an effective off-shell
superpotential, but in the matrix model the $\SH$ is an input parameter
and the effective superpotential emerges from a more direct calculation
we shall perform in~\S3.

Finally, let us fix the overall normalization factor~$\cal C$ in \eq{eq:Zfun}.
Each $\U_\ell$ is integrated over a compact manifold, while each
$\A_\ell^{}=\B_\ell^\dagger$ is integrated over the non-compact $\CC^{\NH\FH}$,
but the integral is Gaussian.
Hence, we let
\be
{\cal C}\ =\ {1\over\left(\mathop{\rm Vol}[{\rm SU}(\NH)]\right)^K}
\times{1\over \eta^{\NH\FH K}}
\label{eq:Cdef}
\ee
where $\eta$ is the Gaussian integral for a single quark mode of
an average mass.
The `average' here does not have to be the arithmetic or the geometric mean,
any representative value will do, and so we use $\gamma\vh$ because
the modes vary in mass from  $\gamma(\vh-\mu_f)$ to $\gamma(\vh+\mu_f)$.
Hence,
\be
\eta\ =\ \int\!\!dx\,d\bar x\,\exp\left(-\NS{}\gamma\vh\,|x|^2\right)\
=\ {2\pi\SH\over\NH\gamma\vh}\,.
\label{eq:eta}
\ee



\subsection{Loop Equations of the Matrix Model}
Having defined our matrix model we now need to verify that it is indeed
dual to the 4D gauge theory which deconstructs the $\rm SQCD_5$.
In this section we shall verify that the loop equations of the matrix
model are similar to the anomaly equations of the gauge theory.
To be precise, the matrix model is dual to a rather small part
of the gauge theory --- namely the subring of its chiral ring involving
either the gaugino condensates or the mesons\footnote{%
   The matrix models also has analogues of the  baryonic and antibaryonic
   operators of the gauge theory, but such ``dual baryons'' are not
   matrices and are rather difficult to handle.
   For this reason, we shall limit our analysis here to the duals of
   the gaugino condensates and mesons only.
   }
--- but the anomaly equations for that part of the gauge theory should be
accurately reproduced by the loop equations of matrix model.

The loop equations follow from infinitesimal holomorphic
 changes of integration variables.
For a simplified example, consider a toy model of a single unitary
$\NH\times\NH$ matrix~$\U$,
\be
{\cal Z}\ = \int_{U(\NH)}\!\!d\omega[\U]\,
\exp\left(-\NS{}\,\tr\Bigl(\WH(\U)\Bigr)\right)
\label{eq:toyZ}
\ee
where $d\omega[\U]$ is the holomorphic form~(\ref{eq:Haar})
of the Haar measure.
Let us change
\be
\U\ \mapsto\ \U'\ =\ \U\ +\ \epsilon\times f(\U)
\label{eq:toyU}
\ee
where $f$ is a holomorphic function of~$\U$.
Generally, this breaks the unitarity of $\U$, hence we should deform
the integration variety from $\Gamma=U(\NH)$ to $\Gamma'$
which spans the $\U'$ for $\U\in U(\NH)$.
However, $\Gamma'\cong\Gamma$ and hence this deformation does not affect
the holomorphic integral~(\ref{eq:toyZ}).
On the other hand, the variable change~(\ref{eq:toyU}) itself has a
non-trivial Jacobian
\be
J\ \equiv\ \frac{d\omega[\U']}{d\omega[\U]}\
=\ 1\ +\,\sum_{i,j}{\delta(\U^{-1}d\U)_{i,j}\over(\U^{-1}d\U)_{i,j}}\
=\ 1\, +\, \epsilon\biggl[ \sum_{jk}{\partial f_{kj}(\U)\over\partial \U_{kj}}
        -\ \NH\tr\Bigl(U^{-1}f(\U)\Bigr)\biggr] .\qquad
\label{eq:toyJ}
\ee
Also, the integrand of the matrix integral changes according to
\be
\exp\left(-\NS{}\,\tr\Bigl(\WH(\U)\Bigr)\right)\
\mapsto\ \mbox{same}\times\left[ 1\
        -\ \epsilon\,\NS{}\,\tr\Bigl(\WH'(\U)f(\U)\Bigr)\right].
\ee
Altogether, we have changed the matrix integral by
\be
\psset{unit=1ex,linecolor=blue,linewidth=1pt,arrowscale=1.8}
\setbox0=\hbox{$\displaystyle{
        \sum_{jk}{\partial f_{kj}(\U)\over\partial \U_{kj}}\
        -\ \NH\tr\Bigl(U^{-1}f(\U)\Bigr)\
        -\ \NS{}\,\tr\Bigl(\WH'(\U)f(\U)\Bigr)
        }$}
\eqalign{
\delta{\cal Z}\ &
=\ \epsilon{\cal Z} \times\left\langle
    \rnode[b]{node:avg}{\psframebox[framearc=1]{\box0}}\right\rangle \cr
&\equiv\ \epsilon\!\!\int_{U(\NH)}\!\!d\omega[\U]\,
    \exp\left(-\NS{}\,\tr\Bigl(\WH(\U)\Bigr)\right)\times
    \Bigl[\;\cnode(0,0.45){0.3}{node:avgd}\;\Bigr] .\cr
}\nccurve[angleA=270,angleB=90]{->}{node:avg}{node:avgd}
\ee
On the other hand, we have done nothing but changed the integration variable
from $\U$ to $\U'$, hence the integral should not change at all.
Therefore, for any holomorphic $\rm matrix\mapsto matrix$ function $f(\U)$
we must have
\be
\vev{ \sum_{jk}{\partial f_{kj}(\U)\over\partial \U_{kj}}\
        -\ \NH\tr\Bigl(U^{-1}f(\U)\Bigr)\
        -\ \NS{}\,\tr\Bigl(\WH'(\U)f(\U)\Bigr)
        }\ =\ 0.
\ee

This toy example shows how to derive loop equations for unitary matrix models.
Let us apply this technology to the unitary $\U_\ell$ matrices of our big
matrix integral~(\ref{eq:Zfun}).
Let us pick one matrix, say $\U_\ell$ and change
\be
U_\ell\ \mapsto\ U_\ell'\ =\ \U_\ell\ +\ \epsilon\times f(\U_\ell,\ldots)
\label{eq:bigU}
\ee
where $f$ is a holomorphic function of the $\U_\ell$ and other matrices of the
model (denoted by the $\ldots$).
To preserve the symmetries of the model, $f(\U_\ell,\ldots)$ must be covariant,
$i.\,e.$ transform like $\U_\ell$ under the $[U(\NH)]^K$.
Also, because the $\U_\ell$ is integrated over the ${\bf S\rm U}(\NH)$ group
manifold rather than ${\rm U}(\NH)$, we must preserve the $\det(\U'_\ell)=
\det(\U_\ell)=1$ condition, hence $f$ must satisfy
\be
\tr\Bigl( \U_\ell^{-1}\,f(\U_\ell,\ldots)\Bigr)\ =\ 0.
\label{eq:bigT}
\ee
On the other hand, we do not need to preserve the unitarity of the $\U'_\ell$
because we may deform the integration variety from $\Gamma=SU(\NH)$ to
a nearby $\Gamma'\cong\Gamma$.
Hence, proceeding exactly as in the toy example above, we find that for
any covariant holomorphic $f$ which satisfies \eq{eq:bigT} we must have
\be
\vev{ \sum_{jk}{\partial f_{kj}(\U_\ell,\ldots)\over\partial (\U_\ell)_{kj}}\
    -\ \NS{}\,\tr\Bigl(\frac{\partial\WH(\U,\A,\B)}{\partial\U_\ell}\,
        f(\U_\ell,\ldots)\Bigr)
    }\
=\ 0.
\label{eq:loopmaster}
\ee

Now let us focus on functions $f$ which depend only on
the unitary link matrices but not on the quark matrices.
By covariance, products of the $\U_{\ell'}$ must be taken in the order
of the quiver, hence $f(\U_\ell,\,\mbox{other}\,\U_{\ell'})$ must be a
linear combination of
$(\U_\ell\U_{\ell-1}\cdots\U_1\U_K\cdots\U_{\ell+1})^p\U_\ell$
for $p=0,1,2,\ldots$, which can be summarized in a power series
\be
\tilde f(X)\ =\,\sum_{p=0}^\infty X^{-1-p}\times
\Bigl(\U_\ell\U_{\ell-1}\cdots\U_1\U_K\cdots\U_{\ell+1}\Bigr)^p\times\U_\ell\
=\ {1\over X\,-\,\U_\ell\U_{\ell-1}\cdots\U_1\U_K\cdots\U_{\ell+1}}\times\U_\ell
\ee
in an auxiliary complex variable~$X$.
In field theory we use this series as it is, but in the matrix model we must
correct for the trace condition~(\ref{eq:bigT}), thus
\be
\psset{unit=1ex,linecolor=blue,linewidth=1pt,arrowscale=1.8}
\setbox0=\hbox{$\displaystyle{
        \;\left[
        {1\over X\,-\,\U_\ell\U_{\ell-1}\cdots\U_1\U_K\cdots\U_{\ell+1}}\
        -\ {1\over\SH}\,R(X)
        \right]\;
        }$}
f(\U_\ell,\ldots)\
=\ \rnode[b]{node:BigBrak}{\psframebox[framearc=1]{\box0}}\
\times\U_\ell
\label{eq:bigF}
\ee
where the second term inside the brackets assures
 $\psset{unit=1ex,linecolor=blue,linewidth=1pt,arrowscale=1.8}
\tr[\;\cnode(0,0.45){0.3}{node:SmallBrak}\;]=0
\nccurve[angleA=270,angleB=90]{->}{node:BigBrak}{node:SmallBrak}$.
Specifically,
\be
R(X)\ \eqdef\ \SN{}\times
\tr\left({1\over X\,-\,\U_\ell\U_{\ell-1}\cdots\U_1\U_K\cdots \U_{\ell+1}}\right)
\label{eq:Rdef}
\ee
which is same for all $\ell$ and remains finite
in the $\NH\to\infty$ limit.
For $f$ as in \eq{eq:bigF}
\be
\label{eq:JacU}
\sum_{jk}{\partial f_{kj}(\U_\ell,\ldots)\over\partial (\U_\ell)_{kj}}\
=\ \NS{2}\left[ XR^2(X)\ -\ \SH R(X)\
        +\ {\SH\over\NH^2}\,\bigl(XR'(X)+R(X)\bigr)\right]
\ee
while
\be
\label{eq:WdefU}
\tr\Bigl(\frac{\partial\WH_{\rm def}}{\partial\U_\ell}\,
        f(\U_\ell,\ldots)\Bigr)\
=\ \NS{}\left[ X\W'(X) R(X)\ -\ \SH\W'(X)\ +\ \SH P(X)\ -\ C R(X)\right]
\qquad\qquad
\ee
where
\bea
\W'(X) &=&\sum_{p=1}^{d}g_p \vh^{pK}X^{p-1},\\
P(X) &=&
{1\over\NH}\,\tr\left((\U_K\cdots\U_1)
        {\W'(X)-\W'(\U_K\cdots\U_1)\over X\,-\,\U_K\cdots\U_1}\right)\\
&&\quad (\mbox{a polynomial of}\ X\ \mbox{of degree}=d-2)\nonumber\\
\mbox{and}\quad C &=&
{1\over\NH}\,\tr\Bigl((\U_K\cdots\U_1)\W'(\U_K\cdots\U_1)\Bigr).
\label{eq:MdefC}
\eea
Note that in the large~$\NH$ limit $C$ and $P(X)$ remain finite, hence
the right hand side of \eq{eq:WdefU} grows like~$\NH$.
By comparison,
\be
\tr\Bigl(\frac{\partial\WH_{\rm hop}}{\partial\U_\ell}\,
        f(\U_\ell,\ldots)\Bigr)\
= O(\FH),
\ee
which is much smaller than $O(\NH)$ in the 't~Hooft limit.
Therefore, \eq{eq:loopmaster} becomes
\be
\vev{ XR^2(X)\ -\ \bigl(X\W'(X)+\SH-C\bigr)\times R(X)\
+\ \SH\W'(X)\ -\ \SH P(X) }\
=\ O(1/\NH)\ \approx\ 0.
\ee
Moreover, for $\NH\to\infty$ matrix averages factorize as
\be
\vev{R^2(X)}\ \to\ \vev{R(X)}^2,\qquad
\vev{C\times R(X)}\ \to \vev{C}\times\vev{R(X)} ,
\ee
and this gives us a loop equation for the $\vev{R(X)}$, namely
\be
{\blue
X\vev{R(X)}^2\ -\ \Bigl( X\W'(X)+\SH-\vev{C}\Bigr)\times\vev{R(X)}\
+\ \SH\W'(X)\ +\ \SH\vev{P(X)}\ =\ 0.
}\label{eq:Rloop}
\ee
\par\goodbreak

On the field theory side, we have a similar quadratic equation
for the gaugino condensate resolvent
\be
R(X)\ =\ \tr\left(
        {(W^\alpha W_\alpha)_\ell\over 32\pi^2}\times
        {1\over X\,-\,\link_\ell\link_{\ell-1}\cdots\link_1
                        \link_K\cdots\link_{\ell+1}}
        \right)
\qquad\qquad(\mbox{same for all}\ \ell).
\label{eq:Rfield}
\ee
In the on-shell chiral ring, this resolvent satisfies
\be
XR^2(X)\ -\ \WW(X)R(X)\ +\ F(X)\ =\ 0
\label{eq:Ranomaly}
\ee
where
\be
\WW(X)\ =\ \sum_{p=1}^dg_p X^p\ +\ \beta v^{N_c}\!\vev{s}
\qquad\qquad(\mbox{same}\ \vev{s_\ell}\ \forall\ell)
\ee
and $F(X)$ is another polynomial (of degree${}<d$) which depends on vacuum
state of the field theory.
Eqs.~(\ref{eq:Rloop}) and~(\ref{eq:Ranomaly}) are obviously dual to each other;
to make them identical we simply need to identify
\be
X\ft\,\longleftrightarrow\,\vh^{K}\times X\mm,\qquad
R(X)\ft\,\longleftrightarrow\,\vh^{-K}\times \vev{R(X)}\mm,\qquad
\label{eq:FMcorr1}
\ee
\be
(\beta v^{N_c}\!\vev{s})\ft\,\longleftrightarrow\,(\SH-\vev{C})\mm,\qquad
F(X)\ft\,\longleftrightarrow\,\vh^{-K}\times
\SH\left(\W'(X)+\vev{P(X)}\right)\mm.
\label{eq:FMcorr2}
\ee
This correspondence explains why $\SH$ is dual
to the net gaugino condensate
\be
S_{\rm net}\ =\ {\vev{\tr(W^\alpha W_\alpha)}\over 32\pi^2}\
=\ \lim_{X\to\infty}(XR(X)).
\ee
Solving the anomaly equation~(\ref{eq:Ranomaly}), we have
\be
R(X)=\ {\WW(X)\,-\,\sqrt{\WW^2(X)-4XF(X)}\over 2X}\
\becomes{X\to\infty}\ {F(X)\over\WW(X)}\,.
\ee
According to \eq{eq:FMcorr2}, the matrix dual of the right hand side is
\be
\vh^{-K}\SH\times{\W'(X)+\vev{(P(X)}\over X\W'(X)+\mbox{const}}\
\becomes{X\to\infty}\ \SH\times{\vh^{-K}\over X\mm}\
=\ {\SH\over X\ft}
\ee
(note that $P(X)$ has lower degree in $X$ than  $\W'(X)$), and therefore
\be
(S_{\rm net})\ft\,\longleftrightarrow\,(\SH)\mm.
\label{eq:FMcorr3}
\ee

\medskip
\centerline{\Large\blue$\star\qquad\star\qquad\star$}
\medskip

The mesonic resolvents of the gauge theory also have matrix duals.
On the field theory side, chiral mesonic operators with quarks and
antiquarks at different quiver nodes are packaged into a bunch of
resolvents
\be
{\cal M}_{\ell',\ell}(X)
=\ \aq_{\ell'}\link_{\ell'-1}\cdots\link_\ell\,
{1\over X\,-\,\link_{\ell-1}\cdots\link_1\link_K\cdots\link_\ell}\,Q_{\ell}
\label{eq:MresF}
\ee
subject to periodicity conditions
\be
-{\cal M}_{\ell'=\ell+K}\ +\ X{\cal M}_{\ell'=\ell}\
=\ M_\ell\ \equiv\ (\aq_\ell Q_\ell)
\label{eq:MperiodF}
\ee
where the right hand side is an ordinary mesonic operator which does
not depend on~$X$.
Besides their quiver indices, the mesonic resolvents are also $N_f\times N_f$
matrices in the flavor space.
On the matrix-model side, they are dual to (the averages of)
$\FH\times\FH$ matrices
\be
{\bf M}_{\ell',\ell}(X)\ =\ \B_{\ell'}\U_{\ell'-1}\cdots\U_{\ell}\,
{1\over X\,-\,\U_{\ell-1}\cdots\U_1\U_K\cdots\U_\ell}\,A_\ell
\label{eq:MresM}
\ee
which satisfy a similar periodicity relation
\be
-{\bf M}_{\ell'=\ell+K}(X)\ +\ X{\bf M}_{\ell'=\ell}(X)\
=\ M_\ell\ \equiv\ (\B_\ell\A_\ell).
\label{eq:MperiodM}
\ee
Because of the $\vh$ factors in the
$\link_\ell\leftrightarrow\U_\ell\times\vh$ correspondence,
in the mesonic sector we expect the gauge-matrix duality to work
according to
\be
X\ft\,\leftrightarrow\,\vh^K \times X\mm,\qquad
{\cal M}_{\ell',\ell}(X)\ft\,\leftrightarrow\,
 \vh^{\ell'-\ell-K}\times\vev{{\bf M}_{\ell',\ell}(X)}\mm ,
\label{eq:FMcorr4}
\ee

To derive loop equations for the mesonic resolvents of the matrix model,
we need to vary the $\A_\ell$ and the $\B_\ell$ matrices independently of
each other.
Note that the conjugacy constraint~$\B_\ell^{}=\A_\ell^\dagger$ is just
a special case of integrating the $2\NH\FH$ complex numbers comprising
each $(\A_\ell,\B_\ell)$ over a variety $\Gamma$ of real dimension $2\NH\FH$,
we ``just happened'' to choose $\Gamma=\{\B_\ell^{}=\A_\ell^\dagger\}$.
But as long as the integrand is a holomorphic function of both $\A_\ell$
and $\B_\ell$ (views as independent variables), we may deform the integration
variety without changing the integral,
\be
\mathop{\int\!\!\!\!\int}\limits_{\B_\ell^{}=\A_\ell^\dagger}\!\!\!
d\A_\ell\,d\B_\ell\,f(\A_\ell,\B_\ell,\ldots)\
=\mathop{\int\!\!\!\!\int}\limits_{\Gamma'}\!
d\A_\ell\,d\B_\ell\,f(\A_\ell,\B_\ell,\ldots)\qquad
\mbox{for}\ \Gamma'\cong\{\B_\ell^{}=\A_\ell^\dagger\}.
\ee
Therefore, {\it small} variations of the $\A_\ell$ and the $\B_\ell$
matrices don't need to be conjugate to each other --- the discrepancy
will deform the integration variety a bit,
but small deformations do not affect the integral.

Minding this rule, let us vary any one $\B_\ell$ matrix while the ``conjugate''
$\A_\ell$ matrix remains unchanged.
Specifically, let
\be
\B_\ell\ \mapsto\ \B'_\ell\ =\ \B_\ell +\ \epsilon\times\B_{\ell'}\times
{\U_{\ell'-1}\cdots\U_\ell\over X\,-\,\U_{\ell-1}\cdots\U_\ell}
\ee
where $\epsilon$ is an infinitesimal $\FH\times\FH$ matrix
in the flavor space.
The Jacobian for such change of variables is
\be
\frac{d^{\FH\NH}\B'_\ell}{d^{\FH\NH}\B_\ell}\
=\ 1\ +\ \delta_{\ell',\ell}\times\tr(\epsilon)\times
\tr\left({1\over X\,-\,\U_{\ell-1}\cdots\U_\ell}\right)\
=\ 1\ +\ \delta_{\ell',\ell}\times\tr(\epsilon)\times
\NS{}\,R(X),\qquad
\ee
while the potential changes by
\be
\delta\WH\ =\ \delta\WH_{\rm hop}\
=\ \gamma\vh\tr\left(\epsilon\times{\bf M}_{\ell',\ell-1}(X)\right)
-\ \gamma\tr\left(\epsilon\times{\bf M}_{\ell',\ell}(X)\times\hat\mu\right).
\ee
But altogether, this is just a change of an integration variable
which does not change the integral at all,
hence
\bea
0 &=&
\vev{\delta(\mbox{Jacobian})}\ -\ \NS{}\,\vev{\delta\WH}\\
&=& \delta_{\ell',\ell}\times\tr(\epsilon)\times\NS{}\,\vev{R(X)}\
-\ \NS{}\,\gamma\tr\left(\epsilon\times\Bigl(
        \vh\vev{{\bf M}_{\ell',\ell-1}(X)}\,
        -\,{\bf M}_{\ell',\ell}(X)\times\hat\mu
        \Bigr)\right)
.\qquad\nonumber
\eea
This must hold true for any $\epsilon$ matrix, therefore
\be
\vh\vev{{\bf M}_{\ell',\ell-1}(X)}\
-\ \vev{{\bf M}_{\ell',\ell}(X)}\times\hat\mu\
=\ \delta_{\ell,\ell'}\times\gamma^{-1}\vev{R(X)}\times
\mbox{\Large\bf1}_{\FH\times\FH}\,.
\label{eq:MloopB}
\ee

Likewise, changing the $\A_{\ell}$ matrix by
\be
\delta\A_\ell\
=\ {\U_{\ell-1}\cdots\U_{\ell'}\over X\,-\,\U_{\ell'-1}\cdots\U_{\ell'}}
\times\A_{\ell'}\times\epsilon
\ee
while the $\B_\ell$ matrix remains unchanged leads to another loop equation,
namely
\be
\vh\vev{{\bf M}_{\ell+1,\ell'}(X)}\
-\ \hat\mu\times\vev{{\bf M}_{\ell,\ell'}(X)}\
=\ \delta_{\ell,\ell'}\times\gamma^{-1}\vev{R(X)}\times
\mbox{\Large\bf1}_{\FH\times\FH}\,.
\label{eq:MloopA}
\ee
Both loop equations~(\ref{eq:MloopB}) and~(\ref{eq:MloopA}) look exactly
like their field theory counterparts:
\be
{\cal M}_{\ell',\ell-1}(X)\ -\ {\cal M}_{\ell',\ell}(X)\times\mu\
=\ {\cal M}_{\ell'+1,\ell}(X)\ -\ \mu\times{\cal M}_{\ell',\ell}(X)\
=\ \delta_{\ell,\ell'}\times\gamma^{-1}\,R(X)\times
\mbox{\Large\bf1}_{N_f\times N_f}\,.
\ee
The only difference is in the $\vh$ factors --- and that is in accordance
with the correspondence rules~(\ref{eq:FMcorr4}) and~(\ref{eq:FMcorr1}).

Having two sets~(\ref{eq:MloopB}) and~(\ref{eq:MloopA}) of mesonic
loop equations is {\sl not} redundant.
Combining both sets with the periodicity equations~(\ref{eq:MperiodM})
allows us to solve for all the mesonic resolvents (or rather their averages)
in terms of a single matrix $\vev{M}$.
The solution works exactly as for the gauge theory, so instead of copying
from~\cite{NaKa04Ch} almost verbatim,
let us simply state the result:
\bea
\label{eq:Meson1}
\vev{{\bf M}_{\ell'=\ell}(X)} &=&
\frac{\vev{M}\,+\,\vh^{-K}\mu^{K-1}\gamma^{-1}\vev{R(X)}\times{\bf1}}
        {X\,-\,\vh^{-K}\hat\mu^K}\,,\\
\label{eq:Meson2}
\vev{{\bf M}_{\ell'>\ell}(X)} &=&
\frac{\hat\mu^{\ell'-\ell-1}}{\vh^{\ell'-\ell}}\times
\frac{\hat\mu\times\vev{M}\,+\,\gamma^{-1}X\vev{R(X)}\times{\bf1}}
        {X\,-\,\vh^{-K}\hat\mu^K}\,,
\eea
where $\vev{M}=\vev{\B_\ell\A_\ell}$ must be same for all~$\ell$
and must commute with the quark mass matrix~$\hat\mu$.\footnote{%
    Strictly speaking, $\vev{M}$ must commute with the $\hat\mu^K$ matrix
    rather than with the $\hat\mu$ itself.
    Also, the matrix elements of $\vev{\B_\ell\A_\ell}$ which commute
    with $\hat\mu^K$ but not with $\hat\mu$ have $\ell$--dependent phases.
    To avoid this mess, we assume that all distinct mass eigenvalues also have
    distinct $K{\rm th\over{}}$ powers.
    }

Finally, there one yet another bunch of loop equations for the mesonic
resolvents stemming from $\A$ and $\B$ dependent variations
of the $\U$ matrices.
Let us pick any one $\U_\ell$ and vary it according to
\be
\psset{unit=1ex,linecolor=blue,linewidth=1pt,arrowscale=1.8}
\setbox0=\hbox{$\displaystyle{
        \A_{\ell+1}\times\epsilon\times\B_\ell\times
        {1\over X\,-\,\U_{\ell-1}\cdots\U_1\U_K\cdots\U_\ell}
        }$}
\delta\U_\ell\
=\ \rnode[b]{node:Uvar}{\psframebox[framearc=1]{\box0}}\
-\ {\tr\Bigl(\U_\ell^{-1}\times(\;\cnode(0,0.45){0.3}{node:Tvar}\;)\Bigr)\over\NH}
\nccurve[angleA=45,angleB=90]{->}{node:Uvar}{node:Tvar}
\times \U_\ell
\label{eq:Dlastloop}
\ee
where the second term assures $\delta\det(\U_\ell)=0$.
The trace in the second term evaluates to
\be
\tr\left(\epsilon\times
        {{\bf M}_{\ell+N,\ell+1}(X)\,-\,{\bf M}_{\ell+N,\ell+1}(0)\over X}
        \right).
\ee
The Jacobian of the variable change~(\ref{eq:Dlastloop}) is given by
\be
\delta J\
=\,\sum_{jk}{\partial(\delta\U_\ell)_{jk}\over\partial(\U_\ell)_{jk}}\
=\ \NS{}\,\tr\left(\epsilon\times\left[\eqalign{
        R(X)&\times{\bf M}_{\ell+K,\ell+1}(X)\cr
        -\ \SH&
        \times{{\bf M}_{\ell+K,\ell+1}(X)\,-\,{\bf M}_{\ell+K,\ell+1}(0)\over X}\cr
       +\ O&(1/\NH^2)\cr
       }\right]\right),
\ee     
while the potential varies according to
\bea
\delta\WH_{\rm hop} &=&
\gamma\vh\tr\Bigl( \epsilon\times
        {\bf M}_{\ell,\ell}(X)\times M_{\ell+1}\Bigr)
   +\ O(1/\NH),\\
\delta\WH_{\rm def} &=&
=\ \tr\left(\epsilon\times\left[\eqalign{
        \W'(X)&\times{\bf M}_{\ell+K,\ell+1}(X)\cr
        -\ C&
        \times{{\bf M}_{\ell+K,\ell+1}(X)\,-\,{\bf M}_{\ell+K,\ell+1}(0)\over X}\cr
        -\ t_{\ell}&(X)\cr
        }\right]\right)
\eea
where $C$ is as in \eq{eq:MdefC} and $t_\ell(X)$ is an $\FH\times\FH$
matrix-valued
polynomial of $X$ whose actual form does not affect the following argument.
Altogether, these changes must cancel out of the matrix integral, hence
demanding
\be
\vev{\delta J}\ -\ \NS\,\vev{\delta\WH}\ =\ 0\qquad\forall\epsilon
\ee
we arrive at
\be
\eqalign{
\gamma\vh\vev{{\bf M}_{\ell,\ell}(X)\times M_{\ell+1}}\ &
-\ \vev{R(X)\times{\bf M}_{\ell+K,\ell+1}(X)}\cr
&=\ \left[\W'(X)\,+\,{\SH-C\over X}\right]
        \times\vev{{\bf M}_{\ell+K,\ell+1}(X)}\
    -\ {\SH-C\over X}\times\vev{{\bf M}_{\ell+K,\ell+1}(0)}\cr
&=\ \vev{t_\ell(X)}\ +\ O(1/\NH).
}\ee
In the large $\NH$ limit matrix averages factorize, and this gives
us another family of loop equations, namely
\be
\eqalign{
\gamma\vh\vev{{\bf M}_{\ell,\ell}(X)}\times\vev{M_{\ell+1}}\ &
+\ \left[\W'(X)\,+\,{\SH-C\over X}\,-\,\vev{R(X)}\right]
        \times\vev{{\bf M}_{\ell+K,\ell+1}(X)}\cr
&-\ {\SH-C\over X}\times\vev{{\bf M}_{\ell+K,\ell+1}(0)}\cr
&=\ \mbox{a polynomial of}\ X.\cr
}\label{eq:Mloop}
\ee

Again, these loop equations are exactly similar to their field theory
counterparts and we may solve them in a similar way.
Making use of \eqrange{eq:Meson1}{eq:Meson2} and \eq{eq:Rloop},
we transform \eq{eq:Mloop} into a quadratic equation for the mesonic
``VEV'' $\vev{M}$, namely
\be
\eqalign{
(\gamma\hat\mu\vev{M})^2\ &
+\ {\hat\mu^K\over X\vh^K}\,[X\W'(X)+\SH-C]\times(\gamma\hat\mu\vev{M})\
    +\ {\hat\mu^K\over X\vh^K}\,F(X)\cr
&=\ \left(1\,-\,{\hat\mu^K\over X\vh^K}\right)\times
    (\mbox{a polynomial of}\ X)
}\label{eq:MloopC}
\ee
Also, the $\vev{M}$ matrix is block-diagonal in the eigenbasis of $\hat\mu$
and does not depend on $X$.
Hence, for each block we may substitute a different value of $X$
into \eq{eq:MloopC} and apply the resulting equation to the block in question;
our choice is $X=\hat\mu_f^K/\vh^K$ which kills the right hand side of
\eq{eq:MloopC} regardless of the matrix-valued polynomial we didn't spell out.
Consequently, each block --- and hence the whole matrix --- satisfies
\be
(\gamma\hat\mu\vev{M})^2\ 
+\ \Bigl((\hat\mu/\vh)^K\W'((\hat\mu/\vh)^K)\,+\,\SH\,-\,C\Bigr)
        \times(\gamma\hat\mu\vev{M})\
+\ F((\hat\mu/\vh)^K)\ =\ 0.
\ee
Note similarity between this equation and the loop equation~(\ref{eq:Rloop});
this allows us to write
\be
\vev{M_\ell}\
=\ -{X\vev{R(X)}^\pm\over\gamma\hat\mu}\quad
\mbox{evaluated for}\ X=(\hat\mu/\vh)^K
\ee
where $\vev{R(X)}^\pm$ on the right hand side indicates the two solutions
of \eq{eq:Rloop}.

This completes our analysis of the loop equations of our matrix model.
Having seen that those equations are dual to the anomaly equations
of the gauge theory of the quiver diagram~(\ref{fig:quiver}),
we can be positive that our model is indeed dual to that gauge theory.

%
%
\section{Calculating the Matrix Integral}
In this section we  evaluate the matrix integral~(\ref{eq:Zfun})
in a sequence of simple steps.
First we integrate over the $\A_\ell$ and $\B_\ell$ matrices dual to
the quarks and the antiquarks.
Second, we reduce the integral over $K$ link matrices $\U_\ell$
to an integral over a single unitary matrix $\M=\U_K\U_{K-1}\cdots\U_1$.
After that we follow the Dijkgraaf--Vafa method adapted to a unitary
rather than hermitian matrix:
we reduce the integral over the whole matrix $\M$ to an integral over
its eigenvalues, and then we use the saddle-point approximation
in the large~$\NH$ limit.
In this limit, the eigenvalue spectrum becomes continuous with density
$\rho(\lambda)$,  the free energy has a $1/\NH$ expansion where
the sphere-level and the disk-level terms  are given
by spectral integrals, --- and we relate the whole shmeer to the loop
equation~(\ref{eq:Rloop}) and the period integrals of its Riemann surface.

Let us integrate over the quark matrices.
The hopping potential~(\ref{eq:WH}) is bilinear with respect to the matrix
elements of the $\A_\ell$ and $\B_\ell$:
\bea
\WH_{\rm hop} &=&
\sum_{\ell,\ell'}\sum_{i,i'}\sum_{f,f'} \B_{\ell',i',f'}^{}\,
        {\cal D}^{\ell',i',f'}_{\quad\ell,i,f}\,\A^{\ell,i,f}_{}\\
\mbox{where}\quad{\cal D}^{\ell',i',f'}_{\quad\ell,i,f} &=&
\vh\delta^{\ell'}_{\,\ell+1}\,(\U_\ell^{})_{\,i}^{i'}\,\delta_{\,f}^{f'}\
-\ \delta^{\ell'}_{\,\ell}\,\delta_{\,i}^{i'}\,\hat\mu_{\,f}^{f'}\,.
\label{eq:Dmatel}
\eea
This makes the quark-matrix integral a Gaussian integral over $K\NH\FH$
independent complex variables, which we evaluate as
\be
\Prl\left\{\mathop{\int\!\!\!\int}\limits_{\B_\ell^{}=\A_\ell^\dagger}
        \!\!d^{\NH\FH}\A_\ell\,d^{\NH\FH}\B_\ell\right\}
\exp\left(-\NS{}\,\WH(\U,\A,\B)\right)\
=\ \left({2\pi\SH\over\NH\gamma}=\eta\vh\right)^{K\NH\FH}\times
{1\over\Det{\cal D}}
\label{eq:matout}
\ee
where the determinant involves all indices: quiver, color, and flavor.
In block form
\be
{\cal D} = \left( \begin{array}{ccccc}
   -\hat\mu & 0 & 0 & \cdots &  \vh\U_K \\
   \vh\U_1 & -\hat\mu & 0 & \cdots & 0 \\
   0 & \vh\U_2 & -\hat\mu & \cdots  & 0 \\
   \vdots & \vdots & \ddots & \ddots & \vdots  \\
   0 & 0 &  \cdots & \vh\U_{K-1} & -\hat\mu \\ 
   \end{array}\right),
\label{eq:Dmat}
\ee
and hence
\be
\Det\limits_{K\NH\FH\times K\NH\FH}({\cal D})
=\ \Det\limits_{\NH\FH\times\NH\FH}
\left(\vh^K\U_K\U_{K-1}\cdots\U_1\,-\,\hat\mu^K\right)\
=\,\prod_{f=1}^{\FH}\det\limits_{\NH\times\NH}
        \left(\vh^K\U_K\U_{K-1}\cdots\U_1\,-\,\hat\mu^K_f\right)
\label{eq:detout}
\ee
where on the right hand side the $\hat\mu_f^{}$
are eigenvalues of the quark mass matrix $\hat\mu$.

Note that the `quark' integral~(\ref{eq:matout}) depends on the link matrices
$\U_\ell$ only through their product $\M\equiv\U_K\U_{K-1}\cdots\U_1$.
Likewise, the deformation superpotential ${\cal W}_{\rm def}$ depends on the
$\U_\ell$ only through their product.
Hence, at this stage, we may write the matrix integral~(\ref{eq:Zfun}) as
\be
{\cal Z}\
=\ {1\over\left(\mathop{\rm Vol}[{\rm SU}(\NH)]\right)^K}\
\Prl\left\{\int_{SU(\NH)}\!\!d\omega[\U_\ell]\right\}\,
f(\M\equiv\U_K\U_{K-1}\cdots\U_1)
\label{eq:SecondInt}
\ee
where
\be
f(\M)\ =\ \exp\left(-\NS{}\WH_{\rm def}(\M)\right)\biggm/
\prod_{f=1}^{\FH}\,\det\limits_{\NH\times\NH}\Bigl(\M-(\mu_f/\vh)^K\Bigr),
\label{eq:fdef}
\ee
and $d\omega[\U_\ell]$ is the Haar measure for $\U_\ell\in SU(\NH)$.
This measure is left-invariant --- for any {\sl fixed} $V\in SU(\NH)$,
$d\omega[V\U]=d\omega[\U]$ --- and this makes it easy to change variables
in the unitary matrix integrals.
In particular, in an integral over $K$ matrices such as~(\ref{eq:SecondInt})
we can set $\U_1=V\U'_1$ (for $\ell=1$ only) and have
\be
d\omega[\U_K]\times\cdots\times d\omega[\U_2]\times d\omega[\U_1=V\U'_1]\
=\ d\omega[\U_K]\times\cdots\times d\omega[\U_2]\times d\omega[\U'_1]\
\ee
for any $V\in SU(\NH)$ which does not depend on the $\U'_1$ matrix, even
if $V$ depends on the {\sl other} unitary matrices $\U_2,\ldots,\U_N$.
Therefore
\begingroup
\interdisplaylinepenalty=1000
\advance\abovedisplayskip by -3pt
\bea
\int_{SU(\NH)}\!\!d\omega[\U_K]\ &\cdots&
\!\!\int_{SU(\NH)}\!\!d\omega[\U_1]\;f(\U_K\cdots\U_1)
\nonumber\\
&=& \!\int_{SU(\NH)}\!\!d\omega[\U_K]\;\cdots\!\!\int_{SU(\NH)}\!\!d\omega[\U_2]
    \int_{SU(\NH)}\!\!d\omega[\U_1']\;f(\U_K\cdots\U_2\times V\U_1')
\nonumber\\
\langle\!\langle\mbox{setting}\ V &=&
(\U_K\cdots\U_2)^{-1}\ \Longrightarrow\
    \U_1'=V^{-1}\U_1=\U_K\cdots\U_2\U_1\equiv\M\,\rangle\!\rangle
\nonumber\\
&=& \!\int_{SU(\NH)}\!\!d\omega[\U_K]\;\cdots\!\!\int_{SU(\NH)}\!\!d\omega[\U_2]
    \int_{SU(\NH)}\!\!d\omega[\U_1'=\M]\,f(\M)
\nonumber\\
\noalign{\vskip 3pt \interdisplaylinepenalty=100\allowbreak}
&=& \Bigl(\mathop{\rm Vol}[{\rm SU}(\NH)]\Bigr)^{K-1}\times
    \int_{SU(\NH)}\!\!d\omega[\M]\,f(\M),
\label{eq:Utriv}
\eea
\endgroup
and hence
\be
{\cal Z}\ =\ {1\over\mathop{\rm Vol}[{\rm SU}(\NH)]}\times
\int_{SU(\NH)}\!\!d\omega[\M]\,f(\M).
\label{eq:OneMatrix}
\ee

Furthermore, according to \eq{eq:fdef}, $f({\cal U})$ depends only on the
eigenvalues of the unitary matrix $\cal U$ but not on its eigenvectors.
Consequently, we decompose the $\NH^2-1$ coordinates of the $SU(\NH)$
group manifold into $\NH-1$ independent eigenvalues $e^{i\lambda_i}$
($i=1,\ldots,\NH$ but $\sum_i\lambda_i\equiv0$ mod~$2\pi$)
and $\NH(\NH-1)$ angular variables
$\theta_{i\neq j}$ describing the eigenvectors.
The Jacobian of this decomposition is given by
\be
\frac{d\omega}{d(\lambda_i,\theta_{i,j})}\
=\ \mathrm{const}\times\prod_{i<j} 4\sin^2{\lambda_i-\lambda_j\over2}
\label{eq:vander}
\ee
 ---  the unitary version of the Vandermonde determinant ---
which depends only on the eigenvalues $\lambda_i$, hence integrating over the
eigenvector variables $\theta_{i,j}$ we arrive at
\be
{\cal Z}\ =\ \frac{\vh^{K\NH\FH}}{\NH!}
\int\limits_0^{2\pi}\!\!{d\lambda_1\over 2\pi}\cdots\!
\int\limits_0^{2\pi}\!\!{d\lambda_{\NH}\over 2\pi}\,
f(\lambda_1,\ldots,\lambda_{\NH})\times
\prod_{i<j} 4\sin^2{\lambda_i-\lambda_j\over2}\times
\sum_{L=-\infty}^{+\infty}e^{iL(\lambda_1+\cdots+\lambda_{\NH})}
\label{eq:eigenint}
\ee
where the last factor is the $\delta$--function for
the $\sum_i\lambda_i$ modulo~$2\pi$.

Thus far we made only exact calculation, but now we turn to approximations
valid in the $\NH\to\infty$ limit.
Re-writing the integrand of \eq{eq:eigenint} in exponential form
\be
\exp\left[\eqalign{
  -\NS{}\sum_{j=1}^{\NH} \W(e^{i\lambda_j})\ &
  -\,\sum_{j=1}^{\NH}\sum_{f=1}^{\FH}\log\Bigl(
        e^{i\lambda_j}-(\mu_f/\vh)^K\Bigr)\cr
  &+\,\sum_{i<j}\log\left(4\sin^2{\lambda_i-\lambda_j\over2}\right)\
    +\ iL\sum_{j=1}^{\NH}\lambda_j\cr
  }\right],
\label{eq:bigexponent}
\ee
we see that all terms in exponent grow with $\NH$.
Hence, in the large $\NH$ limit we may use the saddle-point approximation:
\bea
\log{\cal Z} &\approx&
-\NS{}\sum_{j=1}^{\NH} \W(e^{i\bar\lambda_j})\
    -\,\sum_{j=1}^{\NH}\sum_{f=1}^{\FH}\log\Bigl(
        e^{i\bar\lambda_j}-(\mu_f/\vh)^K\Bigr)\\
&&+\,\sum_{i<j}\log\left(4\sin^2{\bar\lambda_i-\bar\lambda_j\over2}\right)\,
  +\ i\overline{L}\sum_{j=1}^{\NH}\lambda_j\nonumber
\eea
where $(\bar\lambda_1^{},\ldots,\bar\lambda_{\NH}^{})$ and $\overline{L}$
maximize the right hand side of this formula, or rather maximize
its real part and extremize it imaginary part;
generally, this requires moving away from the real axis into
the complex plane.
Or into some other complex space: since the $\lambda_i$ are periodic variables
on the circle~$\RR/2\pi\ZZ$, the $\bar\lambda_i$ move into
the complex cylinder~$\CC/2\pi\ZZ$ rather than into the plane~$\CC$.
As to the $L$, in the large~$\NH$ limit
the maximum happens for $L=O(\NH)$ where the discreteness of $L$ does not
matter any more, hence
$\overline{L}=\NS{}\times\CH$
where $\CH$ is finite and complex.

As usual for matrix models, for $\NH\to\infty$ the spectrum~$\Sigma$ of
$(\bar\lambda_1,\ldots,\bar\lambda_{\NH})$ becomes continuous.
In general, it comprises several continuous line segments
$\Sigma_1,\ldots,\Sigma_n$ on the complex cylinder.
Denoting the spectral density over those lines by
$\NS{}\rho(\lambda)d\lambda$,
we write the free energy of the matrix model as
\be
{\cal F}\
\equiv\ -{\SH^2\over\NH^2}\,\log{\cal Z}\
=\ {\cal F}_S\ +\ {\SH\over\NH}\,{\cal F}_D\ +\ O(1/\NH^2),
\label{eq:freen}
\ee
where
\be
{\cal F}_S\
=\int_\ws\!\!\!d\lambda\,\rho(\lambda)\,
\left[\W(e^{i\lambda})\,-\,i\CH\lambda\right]\
-\ {1\over2}\!\int_\ws\!\!\!d\lambda\,\rho(\lambda)
\!\!\int_\ws\!\!\!d\lambda'\,\rho(\lambda')\,
\log\left(4\sin^2{\lambda-\lambda'\over2}\right)
\label{eq:Sfreen}\\
\ee
and
\be
{\cal F}_D\
=\,\sum_{f=1}^{\FH} \int_\ws\!\!\!d\lambda\,\rho(\lambda)\,
\log\left(e^{i\lambda}-(\mu_f^K/\vh)^K\right).
\label{eq:Dfreen}
\ee
In ``worldsheet'' terms of the $1/\NH$ expansion \cite{Seib02Ad},
the leading contribution
${\cal F}_S$ which comes from ${\cal W}_{\rm def}$ and the Vandermonde determinant
corresponds to genus $g=0$ $i.\,e.$ spherical topology, hence the notation.
Likewise, the quarks' sector contribution corresponds to the disk topology,
hence the notation ${\cal F}_D$.

The spectral density $\rho(\lambda)$ minimizes the free energy
${\cal F}\approx{\cal F_S}$ under constraint
\be
\#(\bar\lambda_i)\,=\,\NH\ \Longrightarrow\
\int_\ws\!\!d\lambda\,\rho(\lambda)\ =\ \SH,
\label{eq:constraintS}
\ee
and this gives us a variational equation
\be
\frac{d}{d\lambda}\,\frac{\delta{\cal F}_S}{\delta\rho(\lambda)}\ =\ 0.
\label{eq:variation}
\ee 
In light of \eq{eq:Sfreen}, this equation becomes
\be
\setbox0=\hbox{$-$}
\setbox1=\hbox to \wd0 {\hss$\displaystyle{\int}$\hss}
\frac{d\W}{d\lambda}\ -\ i\CH\
-\ \mathop{\box0 \kern-\wd1 \box1}\limits_\ws \,
d\lambda'\,\rho(\lambda')\,\cot{\lambda-\lambda'\over2}\ =\ 0
\label{eq:saddle}
\ee
where the bar across the integral sign indicates principal-value integration
over the pole at $\lambda'=\lambda$.
However, it is more convenient to regulate the pole via bypassing it in the
complex plane, so let us introduce the resolvent
\be
\RH(w)\ =\int_\ws\!\!\!d\lambda\,\rho(\lambda)\,
\cot{w-\lambda\over2}
\label{eq:resolv}
\ee
which  has branch cuts along the spectral segments~$\Sigma_1,\ldots,\Sigma_n$
but is analytic elsewhere in the complex cylinder.
In terms of this resolvent, \eq{eq:saddle} applies to the average
of the two sides of any branch cut, hence
\be
\forall\lambda\in\Sigma:\quad
\RH(\lambda\pm i\epsilon)\
=\frac{d\W}{d\lambda}\ -\ i\CH\ \mp\ 2\pi i\rho(\lambda).
\label{eq:dens}
\ee

Together, \eqrange{eq:resolv}{eq:dens} lead to a quadratic equation
for the resolvent $\RH(w)$:
\be
\RH^2(w)\ -\ 2\left[{d\W(e^{iw})\over dw}-i\CH\right]\times\RH(w)\
+\ 2F(e^{iw})\ +\ \SH^2\ =\ 0
\label{eq:lcyl}
\ee
where $F$ is a polynomial function of $e^{iw}$.\footnote{%
  \vrule width 0pt height 20pt
  Integrating $\int\!d\lambda'\rho(\lambda')\int\!d\lambda''\rho(\lambda'')$
  the trigonometric identity
  $$
  \cot{\lambda\pm i\epsilon-\lambda'\over2}\times
    \cot{\lambda'\pm i\epsilon-\lambda''\over2}\,-\,1\
  =\ \cot{\lambda\pm i\epsilon-\lambda''\over2}
  \times\left( \cot{\lambda\pm i\epsilon-\lambda'\over2}\,
    +\,\cot{\lambda'\pm i\epsilon-\lambda''\over2}\right),
  $$
  we arrive after some work at \eq{eq:lcyl} where
  \be
  F(w)\ = \int\limits_\Sigma \!\!d\lambda\,\rho(\lambda)\,
  \cot{w-\lambda\over2}\times\left(
         {d\W(e^{iw})\over dw}\,-\,{d\W(e^{i\lambda})\over d\lambda}\right).
  \ee
  }
Physically, \eq{eq:lcyl} is nothing but the loop equation~(\ref{eq:Rloop})
written in terms of the periodic coordinate $w$ of the complex cylinder
instead of the flat coordinate~$X=e^{iw}$.
Indeed, in the saddle point approximation we have
\be
\vev{R(X)}\ =\ \SN{}\sum_i{1\over X\,-\,\exp(i\bar\lambda_i)}\
=\!\int_\ws\!\!\!d\lambda\,{\rho(\lambda)\over X-e^{i\lambda}}
\ee
and hence
\be
\RH(w)\ =\ 2iX\vev{R(X)}\,-\,i\SH\quad{\rm for}\ X=e^{iw}.
\label{eq:Rmap}
\ee
Likewise
\be
F(X=e^{iw})\ =\ \SH\times\left( \vev{C}\ -\ X\W'(X)\ +\ 2X\vev{P(X)}\right),
\ee
and if we also identify $\CH=\vev{C}$ then \eq{eq:lcyl} becomes
the loop equation~(\ref{eq:Rloop}).

Solving the loop equations gives us
\be
\vev{R(X)}\ =\ {X\W'(X)+\SH-\CH\,-\,Y\over 2X}\
\Longrightarrow\ \RH\ =\ iX\W'(X)\,-\,i\CH\,-\,iY(X)
\label{eq:Rsol}
\ee
where
\be
Y(X)\ =\ \pm
\sqrt{\Bigl(X\W'(X)+\SH-\CH)^2\,-\,4\SH X\Bigl(\W'(X)-\vev{P(X)}\Bigr)}
\label{eq:Ydef}
\ee
has $n\le d={\rm degree}(\W)$ branch cuts which connect simple zeroes
of the discriminant~$Y^2(X)$.
Let us denote such zeroes $a_i^-$ and $a_i^+$ ($i=1,\ldots,n$)\footnote{%
   From now on we use index $i$ to label spectral segments~$\Sigma_i$
   and related parameters rather than individual eigenvalues.
   }
and map them into the $w$ cylinder as $a_i^\pm=\exp(i\theta_i^\pm)$.
Consequently, the $\bar\lambda$ spectrum~$\Sigma$ has precisely $n$
segments~$\Sigma_i$ which begin at $\theta_i^-$ and end at
$\theta_i^+$, the spectral density along each segment is given by
\be
\rho(\lambda)\ =\ {1\over2\pi}\,Y(X=e^{i\lambda})
\label{eq:rhoY}
\ee
({it cf.}\ eqs.~(\ref{eq:dens}) and~(\ref{eq:Rsol})),
and the shapes of the segments follow from $\rho(\lambda)d\lambda$
being real and positive.

{}From the field theory point of view, the segments~$\Sigma_i$ correspond
to confining (or rather pseudo-confining) subgroups $SU(N_i)$ of the diagonal
$SU(N_c)$, and the integrals
\be
\SH_i\
=\int_{\theta_i^-}^{\theta_i^+}\!\!d\lambda\,\rho(\lambda)\
=\ {1\over 4\pi i}\oint\limits_{A_i}\!dw\,\RH(w)\
=\ {1 \over 2\pi i}\oint\limits_{A_i}\!dX \vev{R(X)}
\label{eq:rho}
\ee
are dual to the gaugino condensates
\be
S_i\ =\ \Tr\left.\left(\frac{W^\alpha W_\alpha}{32\pi ^2}\right)\right|_{SU(N_i)}
=\ {1\over 2\pi i}\oint\limits_{A_i}\!dX\,R(X).
\label{eq:contour}
\ee
In both cases the $A_i$ are $A$--cycles of the Riemann surface of the $R(X)\ft
\leftrightarrow\vev{R(X)}\mm$ or their images on the $w$ cylinder;
the $A_i$ surround individual segments of the spectrum as shown below:
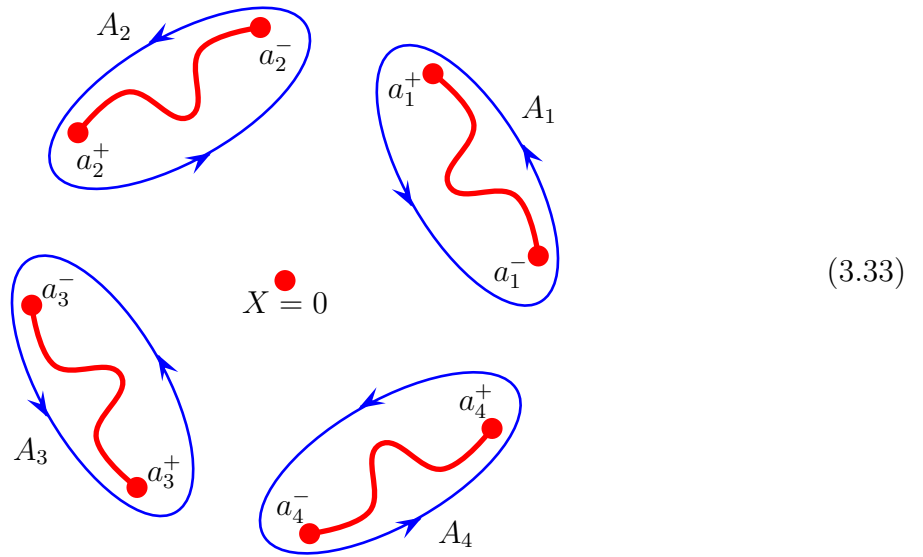
\be
\psset{unit=2.8mm,linewidth=1pt,arrowscale=2}
\def\cut#1{%
    \pscircle*[linecolor=red](-5,11){.5}
    \pscircle*[linecolor=red](+5,11){.5}
    \pscurve[linewidth=2pt,linecolor=red](-5,11)(-2,11.5)(0,9)(+2,11.5)(+5,11)
    \rput[lt](-5,9.5){\rput{*0}{$a^+_{#1}$}}
    \rput[rt](+5,9.5){\rput{*0}{$a^-_{#1}$}}
    \psline[linecolor=blue]{<-}(0,13)(+0.1,13)
    \psline[linecolor=blue]{->}(-0.1,7)(0,7)
    \psellipse[linecolor=blue](0,10)(7,3)
    \rput[b](-1,14.5){\rput{*0}{$A_{#1}$}}
    }
\begin{pspicture}[](-15,-15)(+15,+15)
\rput{300}(0,0){\cut{1}}
\rput {30}(0,0){\cut{2}}
\rput{120}(0,0){\cut{3}}
\rput{210}(0,0){\cut{4}}
\pscircle*[linecolor=red](0,0){0.5}
\rput[t](0,-0.5){$X=0$}
\end{pspicture}
\label{fig:zbc}
\ee

The advantage of the contour integral formalism is that we do not need
to know the exact routes of all the segments but only the general locations
of the $A$--cycles which surround them.
This works for all integrals of the form
$\int\!d\lambda\rho(\lambda)f(\lambda)$, for all kinds of $f(\lambda)$ functions.
For example, \eq{eq:Dfreen} for the disk-level free energy can be written as a sum
of contour integrals
\be
{\cal F}_D\ =\,\sum_{f=1}^{\FH}\frac{1}{4\pi i}\sum_{i=1}^n
\oint\limits_{A_i}\!dw\,\RH(w)\times
\log\left(e^{iw}-(\mu_f/\vh)^K\right).
\label{eq:Df1}
\ee
The sphere-level free energy can also be written as a sum of contour
integrals, but this is more complicated because the logarithm in
the second terms in \eq{eq:Sfreen} has branch cuts of its own.
To disentangle the branch cut structure, let us first rewrite
\eq{eq:Sfreen} as
\be
{\cal F}_S\ =\ {1\over2}\!\int_\ws\!\!\!d\lambda\,\rho(\lambda)\times
\left[ \W(e^{i\lambda})\,-\,i\CH\lambda\,+\,\EH(\lambda)\right]
\label{eq:SFE}
\ee
where
\be
\label{eq:Edef}
\EH(\lambda)\
=\ \W(e^{i\lambda})\,-\,i\CH\lambda\
  -\int_\ws\!\!\!d\lambda'\,\rho(\lambda')\times
    \log\left(4\sin^2{\lambda-\lambda'\over2}\right).
\ee
Naively,
\be
{\delta{\cal F}_S\over\delta \rho(\lambda)}\ =\ \EH(\lambda)
\Longrightarrow\ {d\EH\over d\lambda}\ =\ 0
\ee
({\it cf.}\ \eq{eq:variation}),
but analytic continuation to the complex cylinder yields instead
\be
{d\EH\over d\lambda}\ =\ {d\W\over d\lambda}\ -\ i\CH\
-\ \RH(\lambda)\ =\ iY(X=e^{i\lambda})\ \neq\ 0
\label{eq:derivative}
\ee
and hence the $\EH(\lambda)$ itself is a non-trivial function with branch cuts
along the~$\Sigma_i$ segments.
However, its {\sl average} between the two sides of a cut is locally constant
\be
{\EH(\lambda+i\epsilon)\,+\,\EH(\lambda-i\epsilon)\over2}\
\equiv\ \mathrm{constant}\ \EH_i\quad\mathrm{for}\
\lambda\in\Sigma_i\ \mbox{only}
\label{eq:average}
\ee
because the derivative~(\ref{eq:derivative}) flips sign across the cut.
Therefore, to make sure \eq{eq:variation} is consistent with
the loop equation, we must disambiguate the logarithm in the second term of
\eq{eq:Sfreen} such that
\be
{\bf for}\ \lambda\in\Sigma,\quad
{\delta{\cal F}_S\over\delta \rho(\lambda)}\
=\ {\EH(\lambda+i\epsilon)\,+\,\EH(\lambda-i\epsilon)\over2}\,.
\ee
Hence in light of \eq{eq:average},
\bea
{\cal F}_S &=&
{1\over2}\sum_{i=1}^n\int_{\Sigma_i}\!\!d\lambda\,\rho(\lambda)\times
    \left[ \W(e^{i\lambda})\,-\,i\CH\lambda\,+\,\EH_i\right]
\nonumber\\
&=&{1\over2}\sum_{i=1}^n\left[  \EH_i\times\SH_i\
        +\ {1\over 4\pi i}\oint_{A_i}\!dw\,\RH(w)\times
                \left( \W(e^{iw})\,-\,i\CH w\right)\right].
\label{eq:Sf2}
\eea

The $\EH_i$ can also be calculated as contour integrals.
For general~$\lambda$,
\be
\EH(\lambda)\
=\ \W(e^{i\lambda})\,-\,i\CH\lambda\
-\ \frac{1}{4\pi i}\sum_{i=1}^n\oint\limits_{A_i}\!dw\,\RH(w)
        \times\log\left(4\sin^2{\lambda-w\over2}\right)
\label{eq:Eformula1}
\ee
where the contours $A_i$ should be drawn such as to exclude the branch cuts
of the logarithm.
In particular, the $\lambda$ point itself should be kept outside of all the
$A_i$ contours, and this prevents us from directly evaluating~\eq{eq:Eformula1}
for $\lambda\in\Sigma$.
Instead, for the purpose of calculating an $\EH_i$ in \eq{eq:Sf2} we must first
evaluate the integral~\eq{eq:Eformula1} for $\lambda\not\in\Sigma$,
then take two limits of $\lambda$ approaching
the same point $\lambda_i\in\Sigma_i$ from two opposite sides of the spectrum,
and finally take the average of the two limits.
Alternatively, we may take just one limit of $\lambda$ going to an end point
$\theta_i^+$ or $\theta_i^-$ of the spectral segment ---
at these points the difference between the two sides of the spectrum vanishes
and the averaging becomes unnecessary.
Thus,
\be
\EH_i\ =\ \lim_{\lambda\to\theta_i^\pm}\left[
    \W(e^{i\lambda})\,-\,i\CH\lambda\
     -\ \frac{1}{4\pi i}\sum_{j=1}^n\oint\limits_{A_j}\!dw\,\RH(w)
        \times\log\left(4\sin^2{\lambda-w\over2}\right)
    \right].
\label{eq:Eformula2}
\ee
In the next section, these equations (as well as \eq{eq:Df1}) will help us
calculate the effective superpotential of the matrix model.

We conclude this section with a brief discussion of local {\it versus}
global minima of the free energy.
Eqs.~(\ref{eq:Sf2}) and~(\ref{eq:Eformula2}) give us the 
${\cal F}_S$ minimized with respect to
{\sl local} variations of the spectral density~$\rho(\lambda)$,
hence \eq{eq:rhoY}.
The global minimum requires further minimization with respect to
the free parameters of~\eq{eq:Ydef}, namely  $\CH$ and $n-1$ independent
coefficients of the $\vev{P(X)}$ polynomial.\footnote{%
    $\vev{P(X)}$ has degree $d-2$ and hence $d-1$ coefficients,
    but if we want $n<d$ spectral segments, the polynomial $Y^2(X)$
    ({\it cf.}~\eq{eq:Ydef}) must have $d-n$ double zeroes, which imposes
    $d-n$ constraints on the coefficients of $\vev{P(X)}$.
    Consequently, only $n-1$ of those coefficients may vary independently
    of each other.
    }
Variationally, this implies that $\delta{\cal F}_S/\delta\rho(\lambda)$
should be globally constant over the whole spectrum~$\Sigma$ and not just
individual segments, hence
\be
\EH_1\ =\ \EH_2\ =\ \cdots\ =\ \EH_n\,.
\label{eq:GloMin}
\ee
In addition, we should minimize with respect to the $\CH$ parameter, hence
\be
\frac{\delta{\cal F}_S}{\delta\CH}\
=\!\int_\ws\!\!\!d\lambda\,\rho(\lambda)\times(-i\lambda)\
=\ -{1\over4\pi}\sum_{i=1}^n\oint_{A_i}\!dw\,\RH(w)\times w
\ =\ 0.
\label{eq:Cvar}
\ee

Note that eqs.~(\ref{eq:GloMin}) apply only to the global minimum of the free
energy; in field theory terms this corresponds to taking the gaugino
condensates $S_i$ on-shell.
In the following section, we shall calculate the $W_{\rm eff}$ for the
off-shell $S_i$ or rather $\SH_i$, and this means abandoning the global
minimum and hence eqs.~(\ref{eq:GloMin}).
Instead, eqs.~(\ref{eq:rho}) will determine the coefficients of $\vev{P(X)}$
in terms of the $\SH_i$.

%
%
\section{The Effective Superpotential}
In this section we derive the effective superpotential of our matrix model.
But first, a few general words about effective superpotentials
for the off-shell gaugino condensates in field theory.
In the single--$U(N)$ theory with adjoint matter,
Cachazo {\it et~al.}~\cite{CaDo02Ch}
found that the gauginos of the $U(1)$ center of the $U(N)$ generate auxiliary
supersymmetries of the chiral ring of the gauge theory and hence
\be
W_{\rm eff}(S_1,\ldots,S_n)\
=\,\sum_{i=1}^n N_i {\partial\over\partial S_i}{\cal F}(S_1,\ldots,S_n)\
+\,\sum_{i=1}^n(2\pi i\tau_0+b_i)\times S_i
\ee
where ${\cal F}(S_1,\ldots,S_n)$ is a prepotential of the auxiliary SUSY,
$\tau_0$ is the overall bare gauge coupling, and $b_i$ are integers
distinguishing between specific vacua of the theory.\footnote{%
    By analogy, in the Veneziano--Yankielowicz superpotential
    $W=NS\log S+(2\pi i\tau_0+b)S$, $b$ distinguishes between different vacua
    of the $SU(N)$.
    }
Under gauge-matrix duality, $\tau_0$ is just an arbitrary parameter, but
the prepotential $\cal F$ is dual to the sphere-level
free energy ${\cal F}_S(\SH_1,\ldots,\SH_n)$ of the matrix model.

Adding the quarks --- which couple to the $U(1)$ center --- breaks the auxiliary
supersymmetries, but the subring of the adjoint sector remains supersymmetric.
Hence~\cite{Seib02Ad}, integrating out the mesonic sector of the chiral ring
yields
\be
W_{\rm eff}(S_1,\ldots,S_n)\
=\,\sum_{i=1}^n N_i {\partial\over\partial S_i}{\cal F}(S_1,\ldots,S_n)\
+\,\sum_{i=1}^n(2\pi i\tau_0+b_i)\times S_i\
+\ {\cal F}_Q(S_1,\ldots,S_n)
\label{eq:WeffS}
\ee
where ${\cal F}_Q$ is the quark sector's contribution to the superpotential.
Under gauge-matrix duality, the ${\cal F}_Q$ is dual to the disk-level part
${\cal F}_D$ of the matrix model' free energy.

In our case, the $[SU(N)]^K$ theory does not have a $U(1)$ center(s),
and we cannot promote the gauge symmetry to $[U(N)]^K$ because of
gauge anomaly constraints, but we can add a single $U(1)$ factor
common to all nodes of the quiver diagram~(\ref{fig:quiver}).
Since all the gaugino condensates of our theory belong to subgroups of the
diagonal $SU(N)$, adding the common $U(1)$ effectively promotes
the $SU(N)_{\rm diag}$ to the $U(N)_{\rm diag}$, with similar consequences
for the effective superpotential.
Thus, $W_{\rm eff}(S_1,\ldots,S_n)$ is given by \eq{eq:WeffS} where the
${\cal F}(S_1,\ldots,S_n)$ is dual to the ${\cal F}_S$ of our matrix model
and the ${\cal F}_Q(S_1,\ldots,S_n)$ is dual to its ${\cal F}_D$.

In the matrix model, taking the $\SH_i$ off-shell and evaluating the free
energy as a function ${\cal F}(\SH_1,\ldots,\SH_n)$ means imposing all
$n$ eqs.~(\ref{eq:rho}) as constraints and then minimizing the ${\cal F}
\approx{\cal F}_S$ under those constraints.
Variationally, this implies that $\delta{\cal F}_S/\delta\rho(\lambda)$
should be locally constant along each continuous segment~$\Sigma_i$
of the spectrum but may take different values for different segments.
Therefore, \eq{eq:rhoY} and all the subsequent contour-integral
formul\ae\ of \S3 remain unchanged,
but we throw eqs.~(\ref{eq:GloMin})
out of the window and instead use eqs.~(\ref{eq:rho}) to determine
the coefficients of $\vev{P(X)}$.
%
Consequently, exactly as in~\cite{CaDo02Ch}
\be
\frac{\partial{\cal F}_S}{\partial\SH_i}\
=\ \left.\frac{\delta{\cal F}_S}{\delta\rho(\lambda)}
        \right|_{\lambda\in\Sigma_i}\,
=\ \EH_i\,,
\ee
and hence the effective superpotential of the matrix model is given by
\be
W_{\rm eff}(\SH_1,\ldots,\SH_n)\
=\,\sum_{i=1}^n\NH_i\times\EH_i(\SH_1,\ldots,\SH_n)\
+\,\sum_{i=1}^n(2\pi i\tau_0+b_i)\times \SH_i\
+\ {\cal F}_D(\SH_1,\ldots,\SH_n).
\label{eq:MWeff}
\ee

In the contour integral formalism, the $\EH_i$ in this formula are given
by eqs.~(\ref{eq:Eformula2}).
We can simplify the contour integrals by merging all the $A_i$ cycles
together and then pushing the resulting loop outwards.
The following picture illustrates how this works on the complex cylinder:
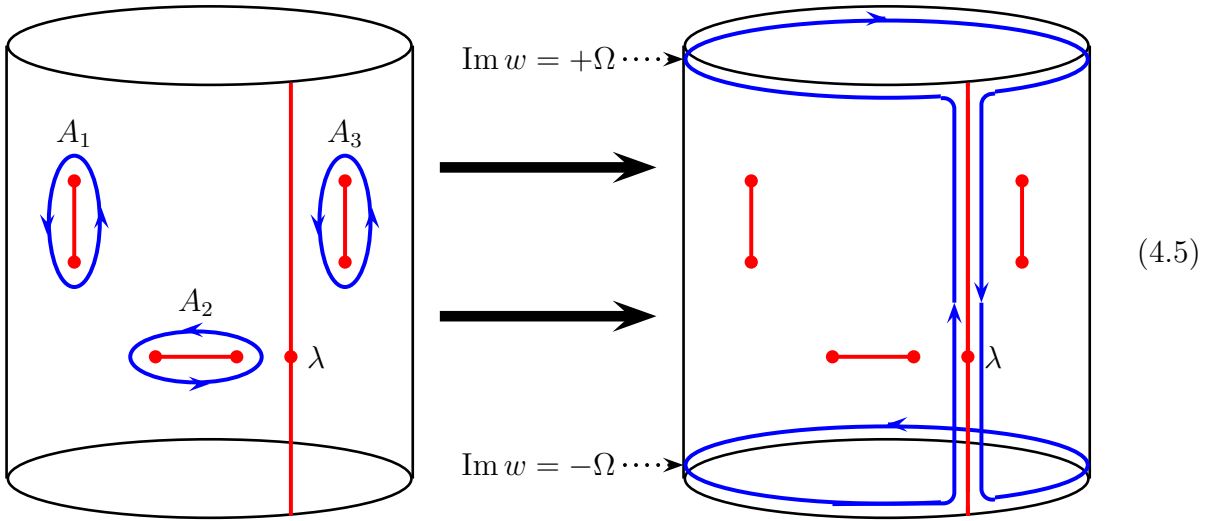
\be
\psset{unit=1.8mm,linewidth=1.5pt,arrowscale=1.25}
\def\eli{%
        \psline[linecolor=blue]{<-}(-1,+2)(0,+2)
        \psline[linecolor=blue]{<-}(+1,-2)(0,-2)
        \psellipse[linecolor=blue](0,0)(5,2.03)
        }
\def\cylinder{%
        \psellipse[linewidth=1pt](0,+16)(15,3)
        \psellipse[linewidth=1pt](0,-16)(15,3)
        \psline[linewidth=1pt](-15,-16)(-15,+16)
        \psline[linewidth=1pt](+15,-16)(+15,+16)
        \psline[linecolor=red,showpoints=true](-10,0)(-10,+6)
        \psline[linecolor=red,showpoints=true](+10,0)(+10,+6)
        \psline[linecolor=red,showpoints=true](-4,-7)(+2,-7)
        \psline[linecolor=red](+6,-18.75)(+6,+13.25)
        \psdot[linecolor=red](+6,-7)
        \rput[l](+7.2,-7){$\lambda$}
        }
\begin{pspicture}[](-41,-19)(+41,+19)
\rput{0}(-25,0){%
    \cylinder
    \rput{90}(-10,+3){\eli}
    \rput{270}(+10,+3){\eli}
    \rput{0}(-1,-7){\eli}
    \rput[b](-1,-4){$A_2$}
    \rput[b](-10,+8.5){$A_1$}
    \rput[b](+10,+8.5){$A_3$}
    }
\rput{0}(+25,0){%
    \cylinder
    \psellipse[linecolor=blue](0,+15)(15,3)
    \psline[linecolor=blue]{->}(-0.1,+18)(0,+18)
    \psframe*[linecolor=white](+4,+12)(+8,+13)
    \psellipse[linecolor=blue](0,-15)(15,3)
    \psline[linecolor=blue]{->}(+0.1,-12)(0,-12)
    \psframe*[linecolor=white](+4,-18)(+8,-17)
    \psline[linecolor=blue,linearc=0.8]{->}(-4,-17.89)(+5,-17.83)(+5,-3)
    \psline[linecolor=blue,linearc=0.8](+5,-3)(+5,+12.17)(+4,12.11)
    \psline[linecolor=blue,linearc=0.8]{->}(+8,+12.46)(+7,+12.35)(+7,-3)
    \psline[linecolor=blue,linearc=0.8](+7,-3)(+7,-17.65)(+8,-17.54)
    \psline[linecolor=red](+6,-18.75)(+6,+13.25)
    }
\rput*[r](+5,+15){$\Im w=+\Omega$}
\psline[linestyle=dotted]{->}(+5,+15)(+10,+15)
\rput*[r](+5,-15){$\Im w=-\Omega$}
\psline[linestyle=dotted]{->}(+5,-15)(+10,-15)
\psline[linewidth=4pt]{->}(-8,+7)(+8,+7)
\psline[linewidth=4pt]{->}(-8,-4)(+8,-4)
\end{pspicture}
\label{fig:FScontour}
\ee
The new contour comprises two vertical lines on two sides of the logarithm's
branch cut, and two loops around the cylinder.
For the loops, we take $\Omega\to\infty$ which gives us
\be
\RH(w)\ \becomes{\Im(w)\,\to\,\pm\infty}\ \mp i\SH
\label{eq:Rapp} 
\ee
and therefore
\bea
\mbox{top loop} &=&
2\pi i\SH\left(\Omega\,-\,\Im(\lambda_i)\right),\\
\mbox{bottom loop} &=&
2\pi i\SH\left(\Omega\,+\,\Im(\lambda_i)\right).
\eea
For the vertical lines, we have discontinuity
\be
\mathop{\rm disc}\left[\log\left(4\sin^2{\lambda-w\over2}\right)\right]\
=\ \cases{+2\pi i & for $\Im w>\Im\lambda$,\cr -2\pi i & $\Im w<\Im\lambda$,\cr }
\ee
and hence
\be
\int_{\rm vert.\atop lines}\!\!dw\,\RH(w)\times\log(\cdots)
=\ 2\pi i\int_\lambda^{+i\Omega}\!\!dw\,\RH(w)\
+\ 2\pi i\int_\lambda^{-i\Omega}\!\!dw\,\RH(w).
\ee
Putting the whole contour integral together, we arrive at
\bea
\EH(\lambda) &=&
\W(e^{i\lambda})\,-\,i\CH\lambda\
    -\ {1\over2}\int_\lambda^{+i\Omega}\!\!dw\,\RH(w)\
    -\ {1\over2}\int_\lambda^{-i\Omega}\!\!dw\,\RH(w)\
    -\ \SH\times\Omega
\label{eq:Sf3}\\
&=& \W(e^{i\lambda})\,-\,i\CH\lambda\
    -\ {1\over2}\int_\lambda^{+i\infty}\!\!dw\,[\RH(w)+i\SH]\
    -\ {1\over2}\int_\lambda^{-i\infty}\!\!dw\,[\RH(w)-i\SH],\qquad
\nonumber
\eea
and at this point, we may take $\lambda=\theta_i^+$ or $\lambda=\theta_i^+$
to calculate the $\EH_i$.

We may also express the $\EH_i$ in term of the $B$--cycle periods of the
Riemann surface of $Y(X=e^{iw})$.
Indeed, plugging \eq{eq:Rsol} into formul\ae~(\ref{eq:Sf3}), we obtain
\bea
\EH_i
&=& {i\over2}\int_{\theta_i^\pm}^{+i\Omega}\!dw\,Y(e^{iw})\
    +\ {i\over2}\int_{\theta_i^\pm}^{-i\Omega}\!dw\,Y(e^{iw})\
    +\ \half\W(e^{iw})\ -\ \SH\Omega
\nonumber\\
&=& -{i\over4}\int_{B_i^+}^{\rm reg}\!dw\,Y(e^{iw})\
    -\ {i\over4}\int_{B_i^-}^{\rm reg}\!dw\,Y(e^{iw})\
    +\ \half\W(e^{iw})\ -\ \SH\Omega ,
\label{eq:Sf4}
\eea
where the $B_i^\pm$ cycle begins at $w=\pm i\infty$ on the physical sheet
of the Riemann surface, crosses the $\Sigma_i$ branch cut to the other sheet,
and then goes back to $w=\pm i\infty$ but on the unphysical sheet,
and the integrals are regularized by starting and stopping at $w=\pm i\Omega$
instead of $w=\pm i\infty$.
The cycles are illustrated on the following figure:
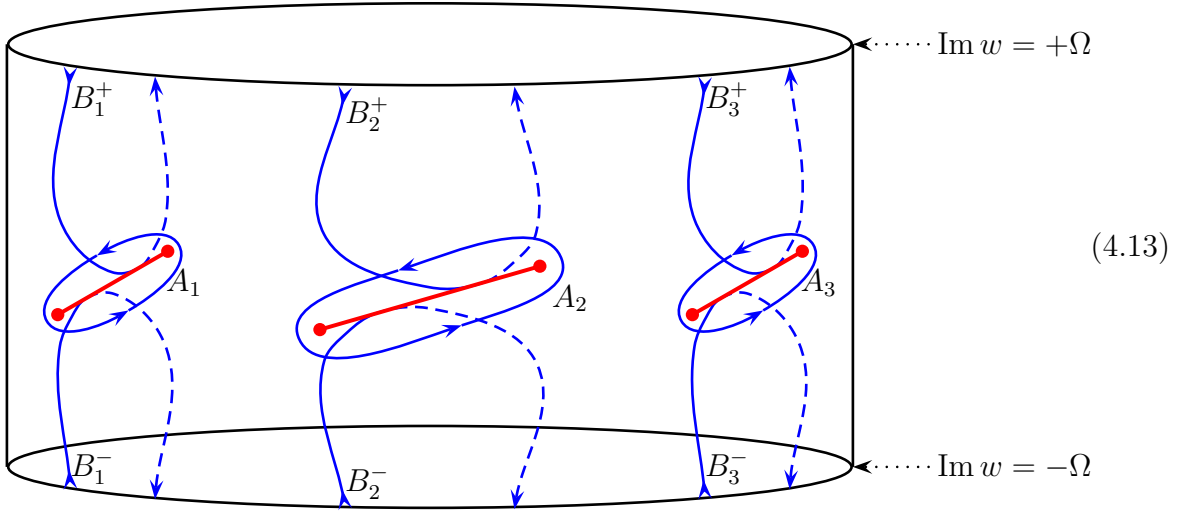
\be
\psset{unit=4pt,linewidth=1pt,arrowscale=1.5}
\def\cut#1(#2){%
    \SpecialCoor
    \rput(0,0){%
        \psecurve[linecolor=blue]{>-}(-3.5,26)(!-4 20 #2 add)%
                (-5,6)(0,1)(+2.6,1.5)(+3.9,3.25)
        \psecurve[linecolor=blue,linestyle=dashed]{->}(0,1)(+2.6,1.5)%
                (+3.9,3.25)(+5,6)(!+4 20 #2 sub)(+3.5,26)
        \rput[lt](-4,19){$B_{#1}^+$}
        }
    \rput{180}(0,0){%
        \psecurve[linecolor=blue,linestyle=dashed]{<-}(-3.5,26)(!-4 20 #2 add)%
                (-5,6)(0,1)(+2.6,1.5)(+3.9,3.25)
        \psecurve[linecolor=blue]{-<}(0,1)(+2.6,1.5)%
                (+3.9,3.25)(+5,6)(!+4 20 #2 sub)(+3.5,26)
        \rput[lb]{*0}(+4,19){$B_{#1}^-$}
        }
    \psline[linecolor=red,showpoints=true,linewidth=1.5pt](-5.2,-3)(+5.2,+3)
    \psecurve[linecolor=blue]{->}(-6.24,-3.6)(+1.5,-2.6)(+6.24,+3.6)(-1.5,+2.6)(-6.24,-3.6)
    \psecurve[linecolor=blue]{->}(+6.24,+3.6)(-1.5,+2.6)(-6.24,-3.6)(+1.5,-2.6)(+6.24,+3.6)
    \rput(+6.5,0){$\,A_{#1}$}
    \NormalCoor
    }
\begin{pspicture}[](-40,-24)(+55,+24)
\psellipse(0,+20)(40,4)
\psellipse(0,-20)(40,4)
\psline(-40,-20)(-40,+20)
\psline(+40,-20)(+40,+20)
\rput{0}(-30,-2.57){\cut1(+0.47)}
\rput{0}(0,-4){\psset{xunit=2}\cut2(0)}
\rput{0}(+30,-2.57){\cut3(-0.47)}
\rput[l](+48,+20){$\Im w=+\Omega$}
\psline[linestyle=dotted]{<-}(+40,+20)(+48,+20)
\rput[l](+48,-20){$\Im w=-\Omega$}
\psline[linestyle=dotted]{<-}(+40,-20)(+48,-20)
\end{pspicture}
\label{fig:Bcontours}
\ee
Note that the each sheet of our Riemann surface is a cylinder with two
distinct infinities --- and that's why we have a double set of $B$--cycles.
Fortunately, most of the extra $B$ cycles are redundant:
\be
\forall i,j:\quad B_i^+\ -\ B_i^-\ \equiv\ B_j^+\ -\ B_j^-\quad
\mbox{modulo}\ A\ \mbox{cycles}.
\ee
and hence
\be
\eqalign{
\EH_i\
=\ \smash{-{i\over2}\int\limits_{B_i^-}^{\rm reg}}\!\!dw\,Y(e^{iw})\ &
+\ \mbox{terms common to all}\ \EH_j\cr
&+\ \mbox{a linear combination of}\ \SH_j\
   \mbox{with integer coefficients.}\cr
}\ee
The last term here reflects the ambiguity of $B$--cycles modulo
$A$~cycles: going from a branch cut $\Sigma_i$ to $\pm i\infty$ one
may choose different passages between the other branch cuts $\Sigma_j$.
In terms of the sphere-level free energy, this corresponds to
different routing of the branch cuts of the
$\log\left(4\sin^2{\lambda-\lambda'\over2}\right)$
around the spectral segments $\Sigma_j$ in both $\lambda$ and $\lambda'$
complex cylinders.
Re-routing the log's branch cuts changes the free energy by
\be
{\cal F}_S(\SH_1,\ldots,\SH_n)\ \to\ {\cal F}_S(\SH_1,\ldots,\SH_n)\
+\ \half\sum_{i,j}c_{ij}\SH_i\SH_j
\ee
for some {\sl integer} coefficients $c_{ij}$, and hence
\be
\EH_i(\SH_1,\ldots,\SH_n)\ \to\ \EH_i(\SH_1,\ldots,\SH_n)\
+\,\sum_j c_{ij}\SH_j\,.
\ee
Fortunately, from the superpotential~(\ref{eq:MWeff}) point of view,
this change amounts to changing the integers~$b_i$.
In other words, we permute the branches belonging to different vacua,
but the overall picture does not change.

Now consider the quark sector's contribution ${\cal F}_D(\SH_1,\ldots,\SH_n)$
to the effective superpotential.
Again, we are going to simplify \eq{eq:Df1} by moving the integration contours,
but it's convenient to do it flavor-by-flavor.
We need to distinguish between massive and massless flavors:
they play different roles in field theory, both in 4D and in 5D\footnote{%
    In deconstruction~\cite{DiKa05Qu}, massive 4D flavors with $\mu_f\approx\vh$
    have light modes and give rise to light flavors in 5D; the massless
    4D flavors do not have no 5D counterparts, but they are needed to adjust
    the Chern--Simons level of the 5D theory.
    From the purely 4D point of view~\cite{NaKa04Ch},
    the massless flavors affect the Coulomb
    branch of the theory but do not give rise to  Higgs branches;
    the on-shell mesonic and baryonic operators include only
    the massive flavors.
    },
and in the matrix model they have different contributions to the disk-level
free energy.
Let us begin with a massive flavor and let~$e^{im_f}=(\hat\mu_f/\vh)^K$.
Then for this flavor we have cuts and contours as shown on the left picture below:
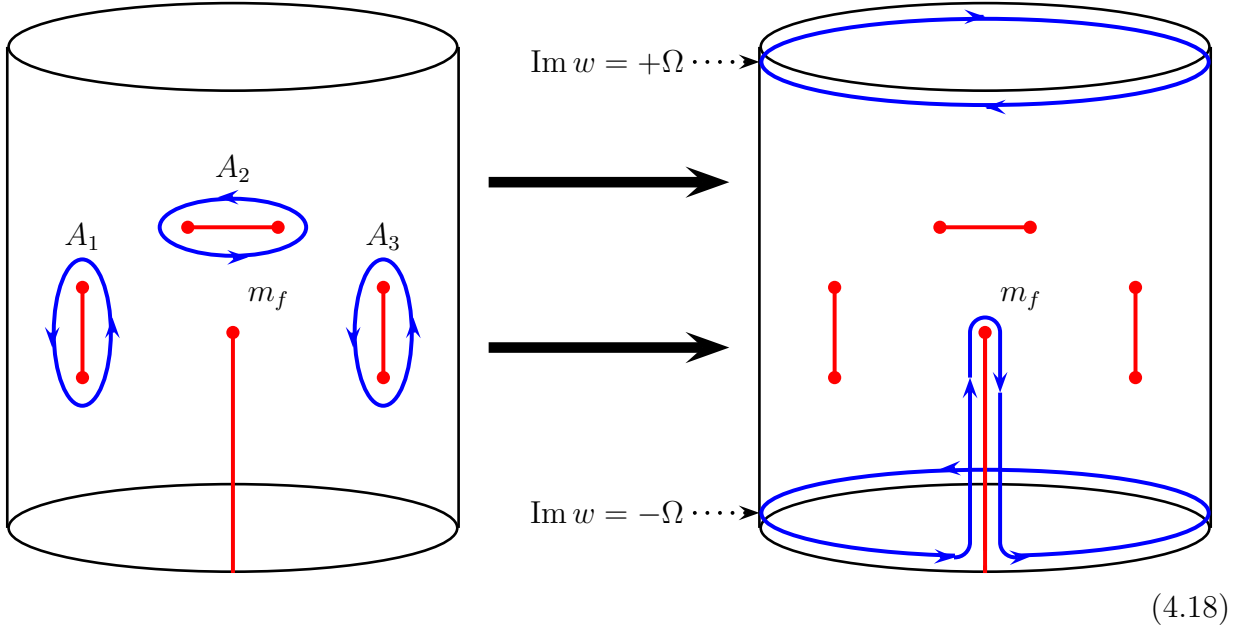
\be
\psset{unit=2mm,linewidth=1.5pt,arrowscale=1.25}
\def\eli{%
        \psline[linecolor=blue]{<-}(-1,+2)(0,+2)
        \psline[linecolor=blue]{<-}(+1,-2)(0,-2)
        \psellipse[linecolor=blue](0,0)(5,2.03)
        }
\def\cylinder{%
        \psellipse[linewidth=1pt](0,+16)(15,3)
        \psellipse[linewidth=1pt](0,-16)(15,3)
        \psline[linewidth=1pt](-15,-16)(-15,+16)
        \psline[linewidth=1pt](+15,-16)(+15,+16)
        \psline[linecolor=red,showpoints=true](-10,0)(-10,-6)
        \psline[linecolor=red,showpoints=true](+10,0)(+10,-6)
        \psline[linecolor=red,showpoints=true](-3,+4)(+3,+4)
        \psline[linecolor=red](0,-3)(0,-19)
        \psdot[linecolor=red](0,-3)
        \rput[bl](1,-1.5){$m_f$}
        }
\begin{pspicture}[](-41,-20)(+41,+19)
\rput{0}(-25,0){%
    \cylinder
    \rput{90}(-10,-3){\eli}
    \rput{270}(+10,-3){\eli}
    \rput{0}(0,+4){\eli}
    \rput[b](0,+7){$A_2$}
    \rput[b](-10,+2.5){$A_1$}
    \rput[b](+10,+2.5){$A_3$}
    }
\rput{0}(+25,0){%
    \cylinder
    \psellipse[linecolor=blue](0,+15)(15,3)
    \psline[linecolor=blue]{->}(-0.1,+18)(0,+18)
    \psline[linecolor=blue]{->}(+0.1,+12)(0,+12)
    \psellipse[linecolor=blue](0,-15)(15,3)
    \psline[linecolor=blue]{->}(-2,-12.04)(-3,-12.09)
    \psframe*[linecolor=white](-2,-18.5)(+2,-17.5)
    \psline[linecolor=blue,linearc=1]{->}(-2,-17.96)(-1,-18)(-1,-6)
    \psline[linecolor=blue,linearc=1]{->}(-1,-6)(-1,-2)(+1,-2)(+1,-7)
    \psline[linecolor=blue,linearc=1](+1,-7)(+1,-18)(+2,-17.96)
    \psline[linecolor=blue]{->}(-3,-17.91)(-2,-17.96)
    \psline[linecolor=blue]{<-}(+3,-17.91)(+2,-17.96)
    \psline[linecolor=red](0,-3)(0,-19)
    }
\rput*[r](+5,+15){$\Im w=+\Omega$}
\psline[linestyle=dotted]{->}(+5,+15)(+10,+15)
\rput*[r](+5,-15){$\Im w=-\Omega$}
\psline[linestyle=dotted]{->}(+5,-15)(+10,-15)
\psline[linewidth=4pt]{->}(-8,+7)(+8,+7)
\psline[linewidth=4pt]{->}(-8,-4)(+8,-4)
\end{pspicture}
\label{fig:FDcontour}
\ee
Again, we merge the $A_i$ contours into a single loop and push it outward
as shown on the right picture above.
This time, we end up with two disconnected contours: a simple loop around
the top of the cylinder, and a loop around the cylinder's bottom
 attached to a vertical loop around the logarithm's cut.
The discontinuity across this cut is simply $+2\pi i$ and hence
\be
\oint_{\rm vert.\atop loop}\!\!dw\,\RH(w)\times
        \log\left(e^{iw}\,-\,e^{im_f}\right)\
=\ 2\pi i\int_{-i\Omega}^{m_f}\!\!dw\,\RH(w).
\ee
And for the loops around the cylinder we again take $\Omega\to\infty$
and use \eq{eq:Rapp}; consequently
\be
\mbox{top loop}\ =\ 2\pi i\SH\times(im_f\,-\,\pi i),\qquad
\mbox{bottom loop}\
=\ 2\pi i\SH\times(\Omega\,+\,\pi i).
\ee
Putting all the loops together, we find that one massive quark flavor
contributes
\be
\left.{\cal F}_D\right|^{{\rm one}\,\mu\neq0}_{\rm flavor}\
=\ {1\over2}\int_{-i\Omega}^{m_f}\!\!dw\,\RH(w)\
+\ {\SH\over2}\times(\Omega+im_f)\
=\ {1\over2}\int_{-i\infty}^{m_f}\!\!dw\,[\RH(w)+i\SH]\ +\ i\SH\times m_f\,.
\qquad\label{eq:Df2}
\ee

For a massless flavor we have
$\log\left( e^{iw}\,-\,(0/\vh)^K\right)\, =\, iw$,
which does not have singularities on the complex cylinder but needs
a branch cut anyway because $w$ is multi-valued.
Consequently, we have contours and cuts as shown below:
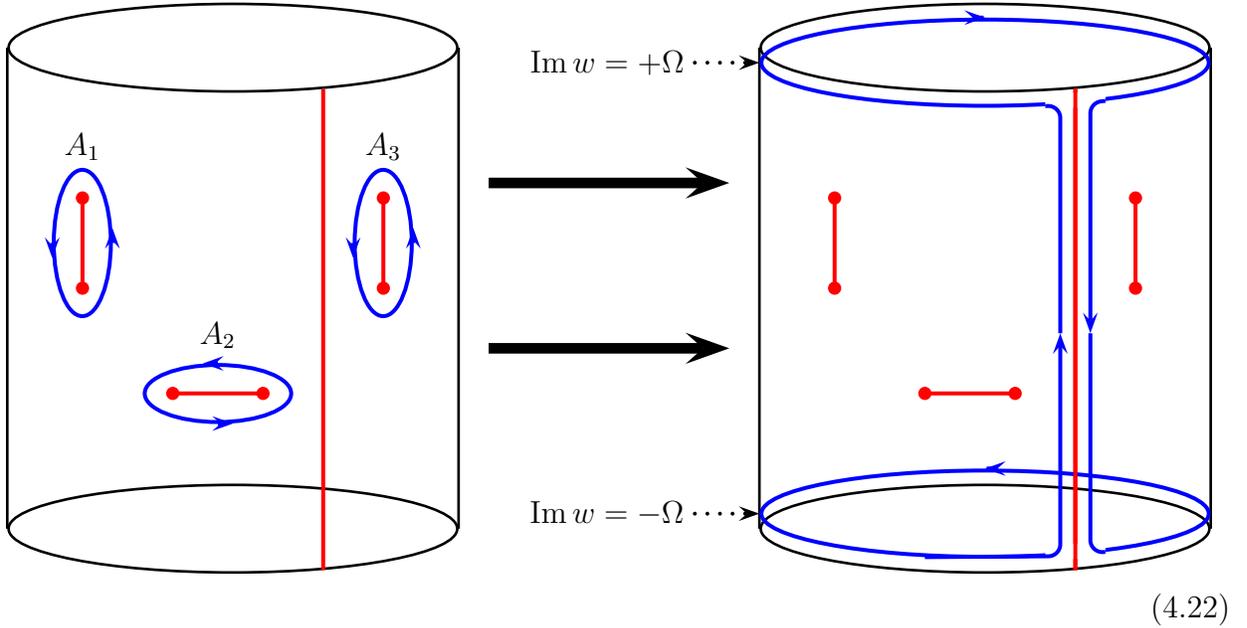
\be
\psset{unit=2mm,linewidth=1.5pt,arrowscale=1.25}
\def\eli{%
        \psline[linecolor=blue]{<-}(-1,+2)(0,+2)
        \psline[linecolor=blue]{<-}(+1,-2)(0,-2)
        \psellipse[linecolor=blue](0,0)(5,2.03)
        }
\def\cylinder{%
        \psellipse[linewidth=1pt](0,+16)(15,3)
        \psellipse[linewidth=1pt](0,-16)(15,3)
        \psline[linewidth=1pt](-15,-16)(-15,+16)
        \psline[linewidth=1pt](+15,-16)(+15,+16)
        \psline[linecolor=red,showpoints=true](-10,0)(-10,+6)
        \psline[linecolor=red,showpoints=true](+10,0)(+10,+6)
        \psline[linecolor=red,showpoints=true](-4,-7)(+2,-7)
        \psline[linecolor=red](+6,-18.75)(+6,+13.25)
        }
\begin{pspicture}[](-41,-20)(+41,+19)
\rput{0}(-25,0){%
    \cylinder
    \rput{90}(-10,+3){\eli}
    \rput{270}(+10,+3){\eli}
    \rput{0}(-1,-7){\eli}
    \rput[b](-1,-4){$A_2$}
    \rput[b](-10,+8.5){$A_1$}
    \rput[b](+10,+8.5){$A_3$}
    }
\rput{0}(+25,0){%
    \cylinder
    \psellipse[linecolor=blue](0,+15)(15,3)
    \psline[linecolor=blue]{->}(-0.1,+18)(0,+18)
    \psframe*[linecolor=white](+4,+12)(+8,+13)
    \psellipse[linecolor=blue](0,-15)(15,3)
    \psline[linecolor=blue]{->}(+0.1,-12)(0,-12)
    \psframe*[linecolor=white](+4,-18)(+8,-17)
    \psline[linecolor=blue,linearc=0.8]{->}(-4,-17.89)(+5,-17.83)(+5,-3)
    \psline[linecolor=blue,linearc=0.8](+5,-3)(+5,+12.17)(+4,12.11)
    \psline[linecolor=blue,linearc=0.8]{->}(+8,+12.46)(+7,+12.35)(+7,-3)
    \psline[linecolor=blue,linearc=0.8](+7,-3)(+7,-17.65)(+8,-17.54)
    \psline[linecolor=red](+6,-18.75)(+6,+13.25)
    }
\rput*[r](+5,+15){$\Im w=+\Omega$}
\psline[linestyle=dotted]{->}(+5,+15)(+10,+15)
\rput*[r](+5,-15){$\Im w=-\Omega$}
\psline[linestyle=dotted]{->}(+5,-15)(+10,-15)
\psline[linewidth=4pt]{->}(-8,+7)(+8,+7)
\psline[linewidth=4pt]{->}(-8,-4)(+8,-4)
\end{pspicture}
\label{fig:FDcontour0}
\ee
Once again we merge the contours into a single loop and push it outward
as shown on the right picture above.
This time we get a connected system of two loops around the cylinder and two
vertical lines, and evaluating the integrals gives us
\bea
\mbox{vertical lines}:&\Longrightarrow&
 2\pi i\int_{-i\Omega}^{+i\Omega}\!\!dw\,\RH(w),\\
\mbox{each horizontal loop}:&\Longrightarrow&
2\pi iS\times(\Omega\,\pm\,\pi i),
\eea
hence altogether one massless flavor yields
\be
\left.{\cal F}_D\right|^{{\rm one}\,\mu=0}_{\rm flavor}\
=\ {1\over2}\int_{-i\infty}^{0}\!\!dw\,[\RH(w)+i\SH]\
+\ {1\over2}\int^{+i\infty}_{0}\!\!dw\,[\RH(w)-i\SH].
\qquad\label{eq:Df3}
\ee

We conclude this section with a complete formula for
the effective superpotential.
Combining eqs.~(\ref{eq:MWeff}), (\ref{eq:Sf3}), (\ref{eq:Df2}),
and (\ref{eq:Df3}) together, we arrive at
\bea
W_{\rm eff} &=&
\sum_{i=1}^n \NH_i\times\left( \W(e^{i\theta_i^+})\,-\,i\CH\theta_i^+\
        +\int_{-i\Omega}^{i\theta_i^+}\!\!dw\,\RH(w)\right)
\nonumber\\
&&+\ {1\over2}\sum_{f=1}^{\FH_1}\int_{-i\Omega}^{m_f}\!\!dw\,\RH(w)\
    -\ {\NH-\FH_2\over2}\times\int_{-i\Omega}^{+i\Omega}\!\!dw\,\RH(w)
\nonumber\\
&&+\ 2\pi i\tau\times\SH_{\rm net}\ +\,\sum_{i=1}^n b_s\SH_i
\label{eq:Wtot}
\eea
where $\FH_1$ is the number of massive flavors, $\FH_2$ is the number
of massless flavors, and $\tau$ is the renormalized gauge coupling
according to
\be
2\pi i\tau\ =\ 2\pi i\tau_0\
+\ {1\over2}\sum_{f=1}^{\FH_1}\log{\mu_f^K\over\vh^K}\
-\ \Omega\times\left(\NH\,-\,\half\FH_1\,-\,\FH_2\right).
\label{eq:tauren}
\ee
Note the coefficient of the cutoff $\Omega$ in this renormalization:
The same combination $N-\half F_1-F_2$ appears in dimensional deconstruction
as the coefficient of the 5D Chern--Simons term.
We are not sure of the physical meaning of this coincidence, but it
probably isn't an accident.

Eq.~(\ref{eq:Wtot}) gives the effective superpotential in terms of the
resolvent $\RH(w)$, and to recast it as a function of the gaugino
condensates $\SH_1,\ldots,\SH_n$ we need to solve eqs.~(\ref{eq:rho})
for the resolvent's parameters.
Since the resolvent with $n$ branch cuts has $n+1$ independent parameters
(including the $\SH=\sum_i\SH_i$), determining all of them requires
one more equation besides $n$ eqs.~(\ref{eq:rho}).
In the matrix model, such $n+1{\rm st\over}$ equation is \eq{eq:Cvar},
which follows from minimizing the free energy with respect to $\CH$.
The contour integrals in \eq{eq:Cvar} are similar to a massless flavor's
contribution to the ${\cal F}_D$ and can be simplified in the same way:
moving the contours according to fig.~(\ref{fig:FDcontour0}), we arrive at
\be
{\partial{\cal F}_S\over \partial\CH}
=\ -{1\over2}\int_{-i\Omega}^{+i\Omega}\!\!dw\,\RH(w)\
-\ \Omega\SH\ =\ 0.
\label{eq:Cvar2}
\ee
The the effective superpotential $W^{\rm eff}(\SH_1,\ldots,\SH_n)$
as a function of the gaugino condensates follows from {\sl combined}
eqs.~(\ref{eq:Cvar2}), (\ref{eq:rho}), and~(\ref{eq:Wtot}).

%
%
\section{Open Questions}
We conclude this paper by discussing open questions raised by
our unitary matrix model.

The first question is specific to the $[SU(N)]^K$ model:
what, if anything, is the gauge-theory dual of \eq{eq:Cvar2}?
The chiral ring of the quiver~(\ref{fig:quiver}) was studied in
detail in \cite{NaKa04Ch}, and none of the anomaly equations there
looks remotely like \eq{eq:Cvar} or \eq{eq:Cvar2}.
Instead, the $n+1{\rm st\over }$ equation for the parameters
of the $R(X)$ resolvent comes from a completely different source,
namely the loop equations for the
link resolvent $T(X)=\tr\left({1\over X-\link_K\cdots\link_1}\right)$.
The analytic properties of the two resolvents $R(X)$ and $T(X)$
require $Y(X)$ ({\it cf.}\ \eq{eq:Ydef}) to have the same branching
points in the $X$ plane as
\be
\sqrt{ \prod_{j=1}^{N_c}(X-\varpi_j)^2\
        -\ 4\left( (-\gamma)^{N_f}\Lambda^{2N_c-N_f}\right)^K
        \times\prod_{f=1}^{N_f}(\mu_f^K-X)\,
        }
\label{eq:QFTSW}
\ee
where $\varpi_1^{}\,\ldots,\varpi_{N_c}^{}$ are the Coulomb moduli
of the $[SU(N)]^K$ theory; only $N_c-1$ of these moduli are independent
because of the ${\bf S\rm U}(N_c)$ constraint
\be
\prod_{j=1}^{N_c}\varpi_j^{}\ =\ \vh^{KN_c}_{}\,.
\label{eq:Coulomb}
\ee
Together, \eqrange{eq:QFTSW}{eq:Coulomb} impose one constraint on
the parameters of the gaugino-condensate resolvent~$R(X)$;
the remaining parameters are related to the individual condensates $S_i$
according to eqs.~(\ref{eq:contour}), exactly as in the matrix model.

Unfortunately, the link resolvent $T(X)$ and the Coulomb moduli
$\varpi_1^{}\,\ldots,\varpi_{N_c}^{}$ of the gauge theory do not
have any matrix duals.
Consequently, \eqrange{eq:QFTSW}{eq:Coulomb} do not make sense on
the matrix-model side of the duality, just like \eq{eq:Cvar2}
does not make any sense on the gauge-theory side.
Ideally, these equations could be dual to each other, but we do not
have any evidence for such duality.
In fact, we do not even know whether \eqrange{eq:QFTSW}{eq:Coulomb}
and~(\ref{eq:Cvar2}) are even mutually consistent.
This remains an open question we hope to answer as soon as possible.

A bigger open question concerns generalization of our unitary
matrix model to other quiver theories.
We believe that the random unitary matrices can be used to model
all kinds of chiral $({\bf n},{\bf\bar n})$ bifundamental fields with
non-zero eigenvalues, but we would like to see how this works
in different models.
Also, it would be interesting to see what exactly goes wrong when
an eigenvalue happens to vanish in some vacuum state of the gauge theory.
We expect the gauge-matrix duality to fail for such vacuum, but we are
not quite sure, and we certainly do not know the specifics (if any)
of this failure.

Finally, we would like to build a matrix model of a
chiral $({\bf n},{\bf\bar n})$ bifundamental with $n\neq m$.
As discussed in the introduction, this calls for complex
$\NH\times\hat M$ matrices integrated over some variety
$\Gamma\subset\CC^{\NH\hat M}$ which has real dimension $\NH\hat M$
and satisfies the symmetry condition~(\ref{eq:symmetry}).
More generally, for a theory with several chiral bifundamentals,
one may use correlated matrices:
instead of independent matrices ${\cal M}_1,\ldots,{\cal M}_K$
each integrated over a separate variety $\Gamma_\ell$ ($\ell=1,\ldots,K$),
the array $({\cal M}_1,\ldots,{\cal M}_K)$ of all the matrices is integrated
over a combined variety $\Gamma\subset\CC^D$ where
$D=\sum_\ell\NH_\ell\hat M_\ell$.
Such combined variety should have real dimension~$D$ (same as the net dimension
of all the $\Gamma_\ell$) and satisfy the generalized symmetry condition
\be
\forall g\in(\mbox{net symmetry group}),\quad
g:\Gamma\mapsto\Gamma'\cong\Gamma .
\label{eq:GenSymm}
\ee
We would like to construct such a variety for an interesting quiver theory,
and then compare the loop equations of the matrix model to the anomaly
equations for the field theory's chiral ring.

\smallskip
\noindent\underline{Acknowledgments}:
We would like to thank Jacob Sonnenschein and Chethan Krishnan for 
interesting conversations on this subject.
V.~K. thanks the HEP department of Tel Aviv University for hospitality
during his multiple visits there.

This material is based upon work supported by the National Science Foundation
under Grant Nos. PHY--0071512 and PHY--0455649,
and with grant support from the US Navy, Office of Naval Research,
Grant Nos. N00014--03--1--0639 and N00014--04--1--0336,
Quantum Optics Initiative.
Neither funding agency should be held responsible for any errors or deficiencies
of this paper.

\bibliography{Bibliografia}
\end{document}